\documentclass[11pt]{amsart}
\usepackage{amsmath,amsfonts,amscd,amssymb,epsf}

\theoremstyle{definition}

\theoremstyle{remark} 
\numberwithin{equation}{section}

\newcommand{\Z}{{\mathbb{Z}}}

\newcommand{\R}{\mathbb{R}}

\newcommand{\pa}{\partial}

\DeclareMathOperator{\sgn}{sgn}

\begin{document}

\title[Casimir effect of Massless fractional Klein-Gordon field]
{Finite Temperature Casimir Effect for a Massless Fractional Klein-Gordon field with Fractional Neumann Conditions}
\author{C.H. Eab$^1$}\email{$^1$Chaihok.E@Chula.ac.th}
\author{S.C. Lim$^2$}\email{$^2$sclim@mmu.edu.my}\author{L.P.
Teo$^{3}$}\email{$^3$lpteo@mmu.edu.my}
\maketitle

\noindent {\scriptsize \hspace{1cm}$^1$Department of Chemistry,
Faculty of Science, Chulalongkorn University,}

\noindent {\scriptsize \hspace{1.1cm} Bangkok 10330, Thailand.}

\noindent {\scriptsize \hspace{1cm}$^2$Faculty of Engineering,
Multimedia University, Jalan Multimedia, }

\noindent {\scriptsize \hspace{1.1cm} Cyberjaya, 63100, Selangor
Darul Ehsan, Malaysia.}

\noindent {\scriptsize \hspace{1cm} $^3$Faculty of Information
Technology, Multimedia University, Jalan Multimedia,}

\noindent{\scriptsize \hspace{1.1cm} Cyberjaya, 63100, Selangor
Darul Ehsan, Malaysia.}

 \begin{abstract}
\vspace{0.5cm} This paper studies the Casimir effect due to
fractional massless Klein-Gordon field confined to parallel plates.
A new kind of boundary condition called fractional Neumann condition
which involves vanishing fractional derivatives of the field is
introduced. The fractional Neumann condition allows the
interpolation of Dirichlet and Neumann conditions imposed on the two
plates. There exists a transition value in the difference between
the orders of the fractional Neumann conditions for which the
Casimir force changes from attractive to repulsive. Low and high
temperature limits of Casimir energy and pressure are obtained. For
sufficiently high temperature, these quantities are dominated by
terms independent of the boundary conditions. Finally, validity of
the temperature inversion symmetry for various boundary conditions
is discussed.

\vspace{0.3cm}

\end{abstract}

\section{\textbf{Introduction}}

Applications of fractional calculus, in particular fractional
differential equations, in transport phenomena in complex and
disordered media have attracted considerable attention during the
past two decades \cite{n1,n2, n3, n4, n5, n6}. However, the use of
fractional calculus in quantum theory is still very limited.
Recently, generalization of quantum mechanics based on fractional
Schrodinger equation has been considered by several authors
\cite{n7, n8, n9,n10, n11}. In quantum field theory, fractional
Klein-Gordon equation \cite{n12, n13, n14, n15, n16} and fractional
Dirac equation \cite{n17, n18} were introduced several years ago,
but further studies on these topics are scarce. It was only lately
that the canonical and stochastic quantization of  fractional
Klein-Gordon field and fractional Maxwell field have been carried
out \cite{n19, n20, n21, n22}.

In this paper, we shall consider another aspect of fractional
Klein-Gordon field, namely, the Casimir effect associated with such
a field. This work is partly motivated by the recent advances in
cosmology, in particular the solid evidences for accelerated
expansion of the universe \cite{n23, n24, n25, n26}, which have
rekindled considerable interest in Casimir effect \cite{n27, n28}.
Casimir energy in extra space-time dimensions \cite{n29} has been
proposed as a possible candidates of dark energy \cite{n30} that is
responsible for the accelerated cosmic expansion. However, in this
paper, we shall not deal directly on the link between Casimir energy
and the dark energy. Instead, we shall study the link between the
possible repulsive nature of the Casimir force and the general
boundary conditions associated with fractional massless Klein-Gordon
field.

In most consideration of Casimir energy between a pair of parallel
plates, the boundary conditions employed are either of Dirichlet
type or Neumann type for both of the plates. A less common pair of
parallel plates has been suggested by Boyer \cite{n31}, with one of
them perfectly conducting and the other infinitely permeable. Boyer
was able to show in the context of random electrodynamics that for
such a set-up the resulting Casimir force is repulsive. It is
possible to show that this unusual pair of plates necessitates mixed
boundary conditions, with the Dirichlet condition for the perfectly
conducting plate, and Neumann condition for the infinitely permeable
plate. Recently, this result has been derived by several authors
using the zeta function method for scalar massless field at zero
temperature \cite{n32} and finite temperature \cite{n33}.

Since this paper studies Casimir effect associated with fractional
Klein-Gordon field, it is not unnatural for one to consider the
fractional generalization of Neumann conditions involving fractional
derivatives. We shall study how repulsive Casimir force due to the
fractional massless Klein-Gordon field can arise under a new type of
boundary conditions, the fractional derivative boundary conditions
(or fractional Neumann conditions). We show that such conditions
allow interpolation between the ordinary Dirichlet and Neumann
conditions.

This paper is organized as follows. In next section we first recall
some basic facts about fractional Klein-Gordon field at zero and
finite temperature. In Section 3, we derive the partition function
and free energy between parallel plates associated with the
fractional massless scalar field at positive temperature using the
generalized thermal zeta function regularization technique. We show
that the Casimir force associated with the massless fractional
scalar field can change from attraction to repulsion as the order of
the fractional Neumann conditions imposed on the parallel plates is
varied. Finally we obtain the low and high temperature limits of
various physical quantities such as free energy and pressure. The
temperature inversion symmetry \cite{n33, n34, n35, n36, n37, LR,
PBPR} will also be discussed.

\section{\textbf{Fractional  Klein--Gordon Field}} In this section, we recall
briefly some basic theory of fractional derivative fields. Let us
consider the Euclidean scalar field $\phi(\mathbf{x}, t)$,
$\mathbf{x}\in\R^D, t\in \R$ with the following Lagrangian
\begin{align}\label{eq13_2}
L=\frac{1}{2}\phi(\mathbf{x},t)\Lambda(-\Delta)\phi(\mathbf{x},t),
\end{align}
where $\Delta=\pa_t^2+\sum_{j=1}^D \pa_j^2$  is the
$(D+1)$--dimensional Euclidean Laplacian operator, and
$\Lambda(-\Delta)$ is a pseudo-differential operator \cite{n38}. In
order to consider $\Lambda(-\Delta)$ of fractional order which
contains the fractional powers $(-\Delta)^{\alpha}$ of Laplacian
operator, we need to define the Riesz fractional derivative and
integral \cite{n39} in order to give these operators a precise
meaning. For a test function in Schwartz space (or a tempered
distribution) $g$, the Fourier transform of $-(\Delta g) (x)$
satisfies $-\widehat{\Delta g}(\xi)=|\xi|^2\hat{g}(\xi)$. This can
be generalized to fractional power of Laplacian operator. For our
purpose, it is sufficient to consider only the real fractional
powers.  For $\alpha\in\R\setminus\{0\}$, and Schwartz functions $g$
we define
\begin{align}\label{eq13_1}
(-\Delta)^{-\alpha/2}g(x)=\left(|\xi|^{\alpha}\hat{f}(\xi)\right)^{\vee}(x)=\begin{cases}
\mathbf{I}^{\alpha}g(x), \hspace{0.5cm}&\alpha > 0,\\
\mathbf{D}^{-\alpha}g(x),&\alpha<0.
\end{cases}
\end{align}
The operators  $ \mathbf{I}^{\alpha}$ and $\mathbf{D}^{\alpha}$
defined in \eqref{eq13_1} for $\alpha>0$ are called respectively the
Riesz fractional integral operator and Riesz fractional differential
operator. We have $ \mathbf{D}^{\alpha}\mathbf{I}^{\alpha}g=g$  and
$\mathbf{I}^{\alpha}\mathbf{I}^{\beta}g=\mathbf{I}^{\alpha+\beta} g$
, $\alpha>0, \beta>0$ for "sufficiently good" functions $g$.

$\Lambda(-\Delta)$ in \eqref{eq13_2} can be expanded in a power
series $\sum_{j}c_j(-\Delta)^j$, and it can be regarded as a
differential operator of infinite order of derivatives. From the
Lagrangian field theory with higher order derivatives \cite{n20,
n40} one gets
\begin{align*}
\sum_{j}(-\Delta)^j\frac{\pa L}{\pa
(-\Delta)^j\phi}=\sum_{j}c_j(-\Delta)_E^j\phi=0,
\end{align*}
and by summing up the series gives the nonlocal field equation
$\Lambda(-\Delta)\phi(\mathbf{x},t) = 0$. Nonlocal field theory with
$\Lambda(-\Delta)=(-\Delta+m^2)^{\alpha}, \alpha>0$ as the
fractional Klein-Gordon operator has been considered by several
authors \cite{n12, n13, n14, n15, n16, n19, n20, n21,n22}. Higher
derivative field theories involving propagator of the form
$(k^2+m^2)^{-n}, n>1$ were first used by Pais and Uhlenbeck
\cite{n41} to obtain a regularized theory without ultraviolet
behavior. Fields with such propagators result in either theories
with ghost states that require a Hilbert space with indefinite
metric, or nonlocal theories without ghost states.

Here we give some remarks on the motivations for introducing
fractional derivative fields. Field theories with nonlocal
Lagrangian of the type \eqref{eq13_2} with nolocality due to kinetic
terms have attracted considerable interest. For examples, nonlocal
kinetic term plays an important role in the (2+1)-dimensional
bosonization \cite{n42, n43}; it also arises in effective field
theories when some degrees of freedom are integrated out in the
underlying local field theory \cite{n44, n45}. One also expects
fractional derivative quantum fields to play an important role in
quantum theories of mesoscopic systems and soft condensed matter
which exhibit fractal character. Such argument can be extended to
quantum field theories in fractal space-time \cite{n46, n47}.

Canonical quantization of nonlocal scalar fractional Klein-Gordon
field has been considered by Amaral and Marino \cite{n19}, and
Barci, Oxman and Rocca \cite{n20}. Free relativistic wave equations
with fractional powers of D'Alembertian operator were studied by
several authors \cite{n13, n14, n15, n16}. Stochastic quantization
of fractional Klein-Gordon  and fractional abelian gauge field has
been considered by Lim and Muniandy \cite{n21}, and finite
temperature fractional Klein-Gordon field is considered in a recent
work \cite{n22}. The two-point Schwinger function of the Euclidean
fractional Klein-Gordon field is given by
\begin{align}\label{eq13_4}
\langle\phi(\mathbf{x},t)\phi(\mathbf{y},s)\rangle
=\frac{1}{(2\pi)^4}\int\limits_{\R^4}\frac{e^{i\mathbf{k}.(\mathbf{x}-\mathbf{y})+ik_4(t-s)}}{(k^2+m^2)^{\alpha}}d^4k.
\end{align} For the Euclidean fractional Klein-Gordon field at finite
temperature  $T=1/\beta$ satisfying the periodic condition $\phi(x,
y, z,0)=\phi(x,y,z,\beta)$ , the two-point Schwinger function
becomes
\begin{align}\label{eq13_5}
\langle\phi(\mathbf{x},t)\phi(\mathbf{y},s)\rangle
=\frac{1}{(2\pi)^3\beta}\sum_{n=-\infty}^{\infty}
\int\limits_{\R^3}\frac{e^{ik_n(x-y)}}{(k_n^2+m^2)^{\alpha}}d^3k,
\end{align}
where $k_n=(\mathbf{k}, \omega_n)$, $\omega_n=2n\pi/\beta$  and
$k_n^2=\mathbf{k}^2+\omega_n^2$. The two point Schwinger functions
for the massless field are given by \eqref{eq13_4} and
\eqref{eq13_5} by putting $m=0$.

In the next section, we shall carry out the computation of Casimir
energy associated with  the massless fractional Klein-Gordon field
confined between two parallel plates imposed with fractional Neumann
boundary conditions. The thermal zeta function technique will be
employed in our calculation. Zeta function method was introduced as
regularization procedure in quantum field theory about two decades
ago \cite{n48, n49, n50}. Basically the zeta function technique
involves three steps. In the case for scalar massless fractional
Klein-Gordon field they are: (I) Determination of  the eigenvalues
$\lambda$ of $(-\Delta)^{\alpha}$
  with appropriate boundary conditions, hence the spectral zeta
function $\zeta_{(-\Delta)^{\alpha}}(s)=\sum_{\lambda}\lambda^{-s}$.
(II) Analytic continuation of the zeta function
$\zeta_{(-\Delta)^{\alpha}}(s)$ to a meromorphic function of the
entire complex plane. (III) Evaluation of $\det(-\Delta)^{\alpha}$
in terms of $ \zeta_{(-\Delta)^{\alpha}}(s)$, that is,
$\det(-\Delta)^{\alpha}=\exp\left(-\zeta_{(-\Delta)^{\alpha}}'(0)\right)$.
For simplicity, the computation will be carried out for scalar
massless fractional Klein-Gordon field. However, one can mimic the
electromagnetic field by the scalar massless field with the two
transverse polarization states of the former taken care of by
multiplying the end results by a factor of two plus some minor
modifications on the possible eigenmodes of the field. In this way,
we can compare our results to some other established results.

\section{\textbf{Free Energy of Massless  Fractional Klein--Gordon Field at Finite
Temperature}}\label{Section3}
 We first assume that the fractional Klein-Gordon field $\phi(\mathbf{x},t)$ is inside a $D$--dimensional space
$\Omega$ which is a rectangular box $ \Omega=[0,
L_1]\times\ldots\times[0, L_{D-1}]\times [0,d]$ such that $d \ll
L_i,$ $1\leq i\leq D-1$. At the end, we let $L_i, 1\leq i\leq D-1$
approach infinity to obtain space between the two hyperplanes
$x_D=0$ and $x_D=d$ in $\R^D$. We want to consider massless
fractional Klein-Gordon field confined in the region $\Omega$ and
maintained in thermal equilibrium at temperature $T=1/\beta$. As
usual \cite{AW}, we impose periodic boundary condition with period
$\beta$ on the imaginary time, i.e.
$$\phi(\mathbf{x}, t)=\phi(\mathbf{x}, t+ \beta),
\hspace{1cm}\forall t\in\R.$$ The Helmholtz free energy of the
system is then given by the equation
\begin{align*} F=-\frac{1}{\beta}\log Z,
\end{align*} where $Z$ is the partition function defined by
\begin{align}\label{eq201}
Z=\int\limits_{\mathcal{BC}}\mathcal{D}[\phi]\exp\left(-\frac{1}{2}\int_0^{\beta}\int_{\Omega}
\phi(\mathbf{x},t)^*(-\Delta)^{\alpha}\phi(\mathbf{x},t)\right)d^D\mathbf{x}dt.
\end{align}Here $\mathcal{BC}$ denotes boundary conditions on the field $\phi$. We impose periodic boundary conditions in
the directions of $x_1, \ldots, x_{D-1}$. In the direction $x_D$, we
can consider different boundary conditions, among them are the
Dirichlet boundary condition with
\begin{align*}
\phi(\tilde{\mathbf{x}}, 0,t)=\phi(\tilde{\mathbf{x}}, d,t)=0,
\hspace{1cm}\forall\;\; \tilde{\mathbf{x}}\in\R^{D-1}, t\in\R,
\end{align*}which corresponds to perfectly conducting plates in the case of electromagnetic field;
the Neumann boundary condition with
\begin{align*}
\left.\frac{\pa}{\pa x_D}\phi(\tilde{\mathbf{x}},
x_D,t)\right|_{x_D=0}=\left.\frac{\pa}{\pa
x_D}\phi(\tilde{\mathbf{x}}, x_D,t)\right|_{x_D=d}=0,
\hspace{1cm}\forall\;\; \tilde{\mathbf{x}}\in\R^{D-1}, t\in\R,
\end{align*}which corresponds to infinitely permeable plates in the case of electromagnetic field;
and the mixed boundary condition with
\begin{align*}
\phi(\tilde{\mathbf{x}}, 0,t)=0,\hspace{0.5cm}\left.\frac{\pa}{\pa
x_D}\phi(\tilde{\mathbf{x}}, x_D,t)\right|_{x_D=d}=0,
\hspace{1cm}\forall\;\; \tilde{\mathbf{x}}\in\R^{D-1}, t\in\R,
\end{align*}which corresponds to Boyer's setup (namely
one plate is perfectly conducting while the other infinitely
permeable) in the case of electromagnetic field. Since we consider
the Casimir effect associated with fractional massless Klein-Gordon
field, one can consider the most general boundary conditions, namely
the fractional boundary conditions
\begin{align}\label{eq1}
\left.\frac{\pa^{\chi}}{\pa
x_D^{\chi}}\phi(\tilde{\mathbf{x}},x_D,t)\right|_{x_D=0}=0,
\hspace{1cm}\left.\frac{\pa^{\mu}}{\pa
x_D^{\mu}}\phi(\tilde{\mathbf{x}},x_D,t)\right|_{x_D=d}=0,
\end{align}where $\chi,\mu\in [0,1]$.  Here we use the  definition of
fractional derivative in terms of Fourier transform:
\begin{align*}
\frac{d^{\eta}f}{d
x^{\eta}}(x)=\frac{1}{2\pi}\int_{-\infty}^{\infty} dk
(ik)^{\eta}e^{ikx}\hat{f}(-k),
\end{align*}where\begin{align*}
(\pm ik)^{\alpha}=|k|^{\alpha}e^{\pm\frac{i\alpha\pi}{2}}\sgn (k)
\end{align*}and $\hat{f}(k)$ is the Fourier transform of $f$.

When $\chi=\mu=0$, one gets the Dirichlet condition for both the
plates. On the other hand, when $\chi=\mu=1$, the boundary
conditions for both plates are that of Neumann type. In the case
with either $\chi=0$, $\mu=1$ or $\chi=1, \mu=0$, we have the Boyer
type boundary condition. For values of $(\chi, \mu)$ other than the
above values, we have fractional Neumann boundary condition for both
plates. One can naively regard such boundary conditions as
correspond to plates which are not perfectly conducting or
infinitely permeable.

Now we want to analyze the condition \eqref{eq1}. If
$$\psi_{k}(z)=Ae^{ikz}+Be^{-ikz}$$are eigen-modes on the $z=x_D$
direction, the requirement \eqref{eq1} is equivalent to
\begin{align*}
A(ik)^{\chi}+&B(-ik)^{\chi}=0, \\
A(ik)^{\mu}e^{ikd}+&B(-ik)^{\mu}e^{-ikd}=0.\nonumber
\end{align*}From these equations, we find
that\begin{align}\label{eq211}B=-e^{i\pi\chi}A,\end{align}
\begin{align}\label{eq3}
2iAk^{\chi+\mu}e^{-\frac{i\pi}{2}(\chi+\mu)}\sin\left(kd-\frac{\pi}{2}(\chi-\mu)\right)=0.
\end{align}From \eqref{eq3}, we find that the value of $k$ has to
be
\begin{align*}
k=\frac{\pi}{d}\left( n +\frac{\chi-\mu}{2} \right),\hspace{1cm}n\in
\Z.
\end{align*}Together with \eqref{eq211},  the eigen-modes in $z$ direction
are given by
\begin{align*}
\psi_n (z) = A
e^{\frac{i\pi}{d}\left(n-\frac{\eta}{2}\right)z}-Ae^{i\pi\chi}e^{-\frac{i\pi}{d}\left(n-\frac{\eta}{2}\right)z},
\end{align*} where $\eta=\mu-\chi$.
We have the following specific cases:

$\chi=\mu=0, \;\;\psi_0=0, \psi_n=-\psi_{-n}; $

$\chi=\mu=1, \;\; \psi_n=\psi_{-n};$

$\chi=0, \mu=1,\;\; \psi_{n}=-\psi_{1-n};$

$\chi=1, \mu=0,\;\; \psi_n=\psi_{-1-n};$

For all other cases, $\psi_n, n\in\Z$ are linearly independent.\\
 We let
$S_{\chi,\mu}=\mathbb{N}$ if $(\chi,\mu)=(0,0)$ or $(0,1)$,
$S_{\chi,\mu}=\mathbb{N}\cup\{0\}$ if $(\chi, \mu)=(1,0)$ or $
(1,1)$ and $S_{\chi,\mu}=\Z$ for all other cases so that
$\{\psi_n(z)\;:\; n\in S_{\chi,\mu}\}$ is a complete set of linearly
independent eigen-modes satisfying the condition \eqref{eq1}. Now it
follows that the eigen-modes of the field $\phi(\mathbf{x}, t)$ are
\begin{align*}
\phi_{\mathbf{k},n,m}(\tilde{\mathbf{x}}, z, t) =
\exp\left(\frac{2\pi i k_1x_1}{L_1}\right)\ldots
\exp\left(\frac{2\pi i
k_{D-1}x_{D-1}}{L_{D-1}}\right)\psi_n(z)\exp\left(\frac{2\pi i
mt}{\beta}\right),
\end{align*}
with $\mathbf{k}=(k_1, \ldots, k_{D-1})\in\Z^{D-1}$, $m\in \Z$,
$n\in S_{\chi,\mu}$.

As is well-known, up to a normalization constant, the path integral
\eqref{eq201} is equal to
\begin{align}\label{eq202}
Z_{\alpha;\chi,\mu}=\left(\prod_{\mathbf{k}\in \Z^{D-1}}\prod_{
m\in\Z}\prod_{n\in
S_{\chi,\mu}}\!'\lambda_{\mathbf{k},n,m}\right)^{-1/2}=\Bigl[\det(-\Delta)^{\alpha}\Bigr]^{-1/2},
\end{align}where
\begin{align*}
\lambda_{\mathbf{k},n,m}=\left( \sum_{j=1}^{D-1}\left(\frac{2\pi
k_j}{L_j}\right)^2+\left(\frac{\pi
}{d}\left(n-\frac{\eta}{2}\right)\right)^2+\left(\frac{2\pi
m}{\beta}\right)^2\right)^{\alpha}.
\end{align*}The prime \;$'$\; in \eqref{eq202} indicates  the omission of  $\lambda_{\mathbf{k}, n,m}=0$ terms. We  compute \eqref{eq202} using zeta
regularization \cite{E1, E2, K}. Namely, we define the spectral zeta
function
\begin{align}\label{eq205}
\zeta_{\alpha;\chi,\mu}(s)=\sum_{\mathbf{k}\in\Z^{D-1}}\sum_{m\in\Z}\sum_{n\in
S_{\chi,\mu}}\!'\lambda_{\mathbf{k},n,m}^{-s}.
\end{align}Then \begin{align*}
\log Z_{\alpha;\chi,\mu}=\frac{1}{2}\zeta_{\alpha; \chi,\mu}'(0).
\end{align*}
Obviously, $\zeta_{\alpha;
\chi,\mu}(s)=\zeta_{\alpha; \mu, \chi}(s)$. 
In terms of the spectral zeta function, the Helmholtz free energy
can be expressed as
\begin{align}
F_{\alpha;\chi,\mu}=-\frac{1}{2\beta}\zeta_{\alpha;\chi,\mu}'(0).
\end{align}We are interested in the limit $L_i
\rightarrow \infty$ for all $1\leq i\leq D-1$. In that case, instead
of the free energy, we consider the free energy density
\begin{align}\label{eq203}
f_{\alpha;\chi,\mu}=\frac{F_{\alpha;\chi,\mu}}{A},\hspace{1cm}\text{where}\;\;A=L_1\ldots
L_{D-1}.
\end{align}As usual, the pressure  is related to the free energy by
the thermodynamic formula
\begin{align}\label{eq35}
P_{\alpha;\chi,\mu}=-\left(\frac{\pa F_{\alpha;\chi,\mu}}{\pa
V}\right)_{T}=-\left(\frac{\pa f_{\alpha;\chi,\mu}}{\pa
d}\right)_{T}.
\end{align}

In order to compute the spectral zeta function
$\zeta_{\alpha;\chi,\mu}(s)$, recall that the Epstein Zeta function
is defined by (see e.g. \cite{E1})
\begin{align}\label{eq11}
Z_E(s; a_1, a_2; \mathbf{g}, \mathbf{h})
=\sum_{\mathbf{n}\in\Z^2}\!'\frac{e^{2\pi i
\mathbf{n}.\mathbf{h}}}{(a_1(n_1+g_1)^2+a_2(n_2+g_2)^2)^s}.
\end{align}Here as usual, the \;$'$\; over the summation means that the term $\mathbf{n}=0$ is omitted when
$\mathbf{g}=0$. The function $Z_E(s; a_1, a_2; \mathbf{g},
\mathbf{h})$ satisfies the functional equation (see e.g. \cite{E1},
page 6)
\begin{align}\label{eq12}
\pi^{-s}\Gamma(s)Z_E(s; a_1, a_2; \mathbf{g},
\mathbf{h})=\frac{e^{-2\pi i
\mathbf{g}.\mathbf{h}}}{\sqrt{a_1a_2}}\pi^{s-1}\Gamma(1-s)Z_E\left(1-s;
\frac{1}{a_1}, \frac{1}{a_2};\mathbf{h},-\mathbf{g}\right).
\end{align}

In the following, we carry out the computation of
$\zeta_{\alpha;\chi,\mu}'(0)$ for various boundary conditions.

 \vspace{0.3cm}\subsection{The case $\chi\neq \mu$ and $(\chi,\mu)\neq
 (0,1),(1,0)$}\label{sec1}~

  \vspace{0.2cm} \noindent This corresponds to the boundary condition
\begin{align*}
&\left.\frac{\pa^{\chi}}{\pa
x_D^{\chi}}\phi(\tilde{\mathbf{x}},x_D,t)\right|_{x_D=0}=0,
\hspace{1cm}\left.\frac{\pa^{\mu}}{\pa
x_D^{\mu}}\phi(\tilde{\mathbf{x}},x_D,t)\right|_{x_D=d}=0,\\
&\hspace{0.8cm}0<\chi <1, \hspace{3.8cm} 0<\mu<1.
\end{align*}
In this case, $\eta\neq 0, \pm 1$. The zeta function
$\zeta_{\alpha;\chi,\mu}$ \eqref{eq205} is given explicitly by
\begin{align*}
\zeta_{\alpha;\chi,\mu}(s)= \sum_{n\in\Z}
\sum_{m\in\Z}\sum_{\mathbf{k}\in \Z^{D-1}}\left(
\sum_{j=1}^{D-1}\left(\frac{2\pi k_j}{L_j}\right)^2+\left(\frac{\pi
}{d}\left(n-\frac{\eta}{2}\right)\right)^2+\left(\frac{2\pi
m}{\beta}\right)^2\right)^{-\alpha s}.
\end{align*}To simplify notation, let
\begin{align*}
a_1=\left(\frac{\pi}{d}\right)^2,\hspace{1cm}a_2=\left(\frac{2\pi}{\beta}\right)^2,\hspace{1cm}c=\frac{\eta}{2}.
\end{align*}As $L_i\rightarrow \infty$ for all $1\leq i\leq D-1$,
\begin{align}\label{eq13}
\zeta_{\alpha;
\chi,\mu}(s)=&\frac{A}{(2\pi)^{D-1}}\sum_{n\in\Z}\sum_{m\in\Z}\int_{\R^{D-1}}d^{D-1}\mathbf{k}
\frac{1}{\left[|\mathbf{k}|^2+a_1(n-c)^2+a_2m^2\right]^{\alpha s}}\\
=&\frac{2\pi^{(D-1)/2}A}{(2\pi)^{D-1}\Gamma\left(\frac{D-1}{2}\right)}\sum_{(n,m)\in\Z^2}
\int_{\R}k^{D-2}dk
\frac{1}{\left[k^2+a_1(n-c)^2+a_2m^2\right]^{\alpha s}}\nonumber\\
=&\frac{2\pi^{(D-1)/2}A}{(2\pi)^{D-1}\Gamma\left(\frac{D-1}{2}\right)}
\int_0^{\infty}k^{D-2}(k^2+1)^{-\alpha s} dk\times\nonumber \\
&\hspace{3cm}\sum_{(n,m)\in\Z^2}\frac{1}{(a_1(n-c)^2+a_2m^2)^{\alpha
s-[(D-1)/2]}}\nonumber\\
=&\frac{A}{(4\pi)^{(D-1)/2}}\frac{\Gamma\left(\alpha
s-\frac{D-1}{2}\right)}{\Gamma(\alpha s)}Z_E\left(\alpha
s-\frac{(D-1)}{2}; a_1, a_2; \mathbf{g},\mathbf{h}\right),\nonumber
\end{align}with
$ \mathbf{g}=(-c,0), \mathbf{h}=\mathbf{0}$. From the functional
equation \eqref{eq12}, we find that
\begin{align}\label{eq15}
\zeta_{\alpha; \chi,\mu}(s)=&\frac{A\pi^{2\alpha
s-D}}{(4\pi)^{(D-1)/2}\sqrt{a_1a_2}}
\frac{\Gamma\left(\frac{D+1}{2}-\alpha s\right)}{\Gamma(\alpha
s)}Z_E\left(\frac{D+1}{2}-\alpha s;
\frac{1}{a_1},\frac{1}{a_2};\mathbf{h};-\mathbf{g}\right)\\=&\frac{A\pi^{2\alpha
s-D}}{(4\pi)^{(D-1)/2}\sqrt{a_1a_2}}\frac{\Gamma\left(\frac{D+1}{2}-\alpha
s\right)}{\Gamma(\alpha s)}\sum_{m\in\Z}\sum_{n\in \Z}\!'
\frac{e^{2\pi i nc}}{([n^2/a_1]+[m^2/a_2])^{[(D+1)/2]-\alpha
s}}.\nonumber
\end{align}This gives
\begin{align*}
\zeta_{\alpha;\chi,\mu}'(0) =&\frac{\alpha
A\;\;\Gamma\left(\frac{D+1}{2}\right)}{2^{D-1}\pi^{(3D-1)/2}\sqrt{a_1a_2}}\sum_{m\in\Z}\sum_{n\in
\Z}\!' \frac{e^{2\pi i nc}}{([n^2/a_1]+[m^2/a_2])^{(D+1)/2}}\\=&
\frac{\alpha
A\;\;\Gamma\left(\frac{D+1}{2}\right)}{2^{D-1}\pi^{(3D-1)/2}\sqrt{a_1a_2}}\Biggl(2a_2^{(D+1)/2}\zeta_R(D+1)\\
&+2a_1^{(D+1)/2}\sum_{n=1}^{\infty} \frac{\cos(2\pi n c)}{n^{D+1}}
+4\sum_{m=1}^{\infty}\sum_{n=1}^{\infty} \frac{\cos(2\pi n
c)}{([n^2/a_1]+[m^2/a_2])^{(D+1)/2}}\nonumber\Biggr),
\end{align*}where $\zeta_R(s)$ is the Riemann zeta function. Define
\begin{align*}
\xi=\frac{d}{\pi\beta}=\frac{1}{2\pi}\sqrt{\frac{a_2}{a_1}}.
\end{align*}In terms of $\xi$, the free energy density \eqref{eq203}
is equal to
\begin{align}\label{eq207}
f_{\alpha;
\chi,\mu}=&-\frac{2\alpha\Gamma\left(\frac{D+1}{2}\right)\pi^{\frac{D+1}{2}}}{d^D}\Biggl(
\xi^{D+1}\zeta_R(D+1)+\frac{1}{(2 \pi)^{D+1}}\sum_{n=1}^{\infty}
\frac{\cos(\pi n \eta)}{n^{D+1}}\\
&\hspace{3cm}+2\xi^{D+1}\sum_{m=1}^{\infty}\sum_{n=1}^{\infty}
\frac{\cos(\pi n \eta)}{(m^2+(2\pi \xi
n)^2)^{(D+1)/2}}\nonumber\Biggr),
\end{align}and the pressure \eqref{eq35} is given by
\begin{align}\label{eq208}
P_{\alpha;\chi,\mu}=&\frac{2\alpha\Gamma\left(\frac{D+1}{2}\right)\pi^{\frac{D+1}{2}}}{d^{D+1}}\Biggl(
\xi^{D+1}\zeta_R(D+1)-\frac{D}{(2 \pi)^{D+1}}\sum_{n=1}^{\infty}
\frac{\cos(\pi n \eta)}{n^{D+1}}\\
&\hspace{3cm}+2\xi^{D+1}\sum_{m=1}^{\infty}\sum_{n=1}^{\infty}
\frac{\cos(\pi n \eta)(m^2-D(2\pi \xi n)^2)}{(m^2+(2\pi \xi
n)^2)^{(D+3)/2}}\nonumber\Biggr).
\end{align}By using the formula \textbf{9.622} in \cite{GR},
\begin{align*}
B_{2n}(x)=\frac{(-1)^{n-1}2(2n)!}{(2\pi)^{2n}}\sum_{k=1}^{\infty}\frac{\cos(2\pi
kx)}{k^{2n}},
\end{align*}where $B_{2n}$ is the Bernoulli polynomial of order $2n$. In particular, when $D=3$, using $B_4(x)=x^4-2x^3+x^2-1/(30)$
and $\zeta_R(4)=-\pi^4B_4(0)/3=\pi^4/90$, we find that the free
energy density and the pressure are given respectively by
\begin{align*}
 f_{\alpha; \chi,\mu} =-\frac{\alpha}{d^3} \left(
\frac{\pi^6\xi^4}{45}-\frac{\pi^2}{24
}B_4\left(\frac{\eta}{2}\right)+4\pi^2\xi^4\sum_{m=1}^{\infty}\sum_{n=1}^{\infty}
\frac{\cos(\pi n \eta)}{(m^2+(2\pi\xi n)^2)^{2}} \right),
\end{align*}
\begin{align*}
P_{\alpha;\chi,\mu}=\frac{\alpha }{d^4}\left(
\frac{\pi^6\xi^4}{45}+\frac{\pi^2}{8
}B_4\left(\frac{\eta}{2}\right)+4\pi^2\xi^4\sum_{m=1}^{\infty}\sum_{n=1}^{\infty}
\frac{\cos(\pi n \eta)(m^2-3(2\pi\xi n)^2)}{(m^2+(2\pi\xi n)^2)^{3}}
\right).
\end{align*}

\vspace{0.3cm}
\subsection{ The case $\chi=\mu=0$ [Dirichlet Boundary Condition]}\label{sec2} ~

 \vspace{0.2cm}\noindent
In this case, the zeta function $\zeta_{\alpha;0,0}$ \eqref{eq205}
is given by\begin{align}\label{eq7} \zeta_{\alpha;0,0}(s)=
\sum_{n=1}^{\infty}\sum_{m\in\Z}\sum_{\mathbf{k}\in \Z^{D-1}}\left(
\sum_{j=1}^{D-1}\left(\frac{2\pi k_j}{L_j}\right)^2+\left(\frac{\pi
n}{d}\right)^2+\left(\frac{2\pi m}{\beta}\right)^2\right)^{-\alpha
s}.
\end{align} Using the same method as in Section \ref{sec1}, we
obtain\begin{align}\label{eq20} \zeta_{\alpha; 0,0}(s)
=&\frac{A}{(4\pi)^{(D-1)/2}}\frac{\Gamma\left(\alpha
s-\frac{D-1}{2}\right)}{\Gamma(\alpha
s)}\sum_{n=1}^{\infty}\sum_{m\in \Z}\frac{1}{(a_1n^2+a_2m^2)^{\alpha
s-[(D-1)/2]}}\\=&\frac{A\Gamma\left(\alpha s-\frac{D-1}{2}\right)
}{2(4\pi)^{(D-1)/2}\Gamma(\alpha s)}\Biggl(Z_E\left(\alpha
s-\frac{D-1}{2}; a_1, a_2\right)\nonumber\\
&\hspace{5cm}-2a_2^{[(D-1)/2]-\alpha s}\zeta_R(2\alpha
s-(D-1))\Biggr)\nonumber\\
=&\frac{A}{2(4\pi)^{(D-1)/2}\Gamma(\alpha
s)}\Biggl(\frac{\pi^{2\alpha s-D}\Gamma\left(\frac{D+1}{2}-\alpha
s\right)}{\sqrt{a_1 a_2}}Z_E\left(\frac{D+1}{2}-\alpha
s;\frac{1}{a_1},\frac{1}{a_2}; \mathbf{0},\mathbf{0}\right)\nonumber\\
&\hspace{1cm}-2a_2^{(D-1)/2-\alpha s}\pi^{2\alpha
s-D+(1/2)}\Gamma\left(\frac{D}{2}-\alpha s\right)\zeta_R(D-2\alpha
s)\Biggr)\nonumber.\end{align}Here we have used the functional
equation
\begin{align*}
\pi^{-\frac{s}{2}}\Gamma\left(\frac{s}{2}\right)\zeta_R(s)=\pi
^{\frac{s-1}{2}}\Gamma\left(\frac{1-s}{2}\right)\zeta_R(1-s)\end{align*}for
Riemann zeta function. The first term in \eqref{eq20} is half of the
term \eqref{eq15} with $\eta=0$. Therefore, we find that the free
energy density and the pressure are given respectively by
\begin{align}\label{eq14_1}
f_{\alpha;
0,0}=&-\frac{\alpha\Gamma\left(\frac{D+1}{2}\right)\pi^{\frac{D+1}{2}}}{d^D}\Biggl(
\xi^{D+1}\zeta_R(D+1)+\frac{\zeta_R(D+1)}{(2
\pi)^{D+1}}-\frac{\Gamma\left(\frac{D}{2}\right)}
{2\sqrt{\pi}\Gamma\left(\frac{D+1}{2}\right)}\xi^D\zeta_R(D)\\
&\hspace{3cm}+2\xi^{D+1}\sum_{m=1}^{\infty}\sum_{n=1}^{\infty}
\frac{1}{(m^2+(2\pi \xi n)^2)^{(D+1)/2}}\nonumber\Biggr),\nonumber
\end{align}
\begin{align*}
P_{\alpha;0,0}=&\frac{\alpha\Gamma\left(\frac{D+1}{2}\right)\pi^{\frac{D+1}{2}}}{d^{D+1}}\Biggl(
\xi^{D+1}\zeta_R(D+1)-\frac{D\zeta_R(D+1)}{(2
\pi)^{D+1}}\\
&\hspace{3cm}+2\xi^{D+1}\sum_{m=1}^{\infty}\sum_{n=1}^{\infty}
\frac{\cos(\pi n \eta)(m^2-D(2\pi \xi n)^2)}{(m^2+(2\pi \xi
n)^2)^{(D+3)/2}}\nonumber\Biggr).
\end{align*}In particular, when $D=3$,
\begin{align*}
f_{\alpha; 0, 0}=-\frac{\alpha}{2d^3} \left(
\frac{\pi^6\xi^4}{45}+\frac{\pi^2}{720
}+4\pi^2\xi^4\sum_{m=1}^{\infty}\sum_{n=1}^{\infty}
\frac{1}{(m^2+(2\pi\xi n)^2)^{2}}
-\frac{\pi^2\xi^3}{2}\zeta_R(3)\right),
\end{align*}and
\begin{align*}
P_{\alpha; 0, 0}=\frac{\alpha}{2d^4}\left(
\frac{\pi^6\xi^4}{45}-\frac{\pi^2}{240
}+4\pi^2\xi^4\sum_{m=1}^{\infty}\sum_{n=1}^{\infty} \frac{m^2-3(2\pi
\xi n)^2}{(m^2+(2\pi\xi n)^2)^{3}} \right).
\end{align*}

 \vspace{0.3cm}
\subsection{The case $\chi=\mu\neq 0, 1$}\label{sec4}~

\vspace{0.2cm}\noindent This corresponds to the boundary condition
\begin{align*}
&\left.\frac{\pa^{\chi}}{\pa
x_D^{\chi}}\phi(\tilde{\mathbf{x}},x_D,t)\right|_{x_D=0}=\left.\frac{\pa^{\chi}}{\pa
x_D^{\chi}}\phi(\tilde{\mathbf{x}},x_D,t)\right|_{x_D=d}=0,\hspace{0.8cm}0<\chi
<1.
\end{align*}

 In this case, $\eta=0$
and the associated zeta function \eqref{eq205} becomes
\begin{align}\label{eq298} \zeta_{\alpha;\chi,\chi}(s)=&
\sum_{(\mathbf{k}, n,m)\in \Z^{D+1}\setminus\{\mathbf{0}\}}\left(
\sum_{j=1}^{D-1}\left(\frac{2\pi k_j}{L_j}\right)^2+\left(\frac{\pi
n }{d}\right)^2+\left(\frac{2\pi m}{\beta}\right)^2\right)^{-\alpha
s},\end{align} which can be written as the sum of two terms\begin{align*}
\zeta_{\alpha;\chi,\chi}(s)=& \sum_{\mathbf{k}\in
\Z^{D-1}}\sum_{(n,m)\in\Z^2\setminus\{(0,0)\}}\left(
\sum_{j=1}^{D-1}\left(\frac{2\pi k_j}{L_j}\right)^2+\left(\frac{\pi
n }{d}\right)^2+\left(\frac{2\pi m}{\beta}\right)^2\right)^{-\alpha
s}\\
&+\sum_{\mathbf{k}\in \Z^{D-1}\setminus\{\mathbf{0}\}}\left(
\sum_{j=1}^{D-1}\left(\frac{2\pi k_j}{L_j}\right)^2\right)^{-\alpha
s}\nonumber =
\zeta_{\alpha;\chi,\chi}^1(s)+\zeta_{\alpha;\chi,\chi}^2(s).
\end{align*}The first term $\zeta_{\alpha;\chi,\chi}^1(s)$ can be computed as in Section \ref{sec1} and the result is the same
as \eqref{eq15} with $\eta=0$. For the second term
$\zeta_{\alpha;\chi,\chi}^2(s)$, we want to verify in the following
that it does not contribute to the free energy density. We have
\begin{align*}
\zeta_{\alpha;\chi,\chi}^2(s)=&2\sum_{(k_1,\ldots, k_{D-2})\in
\Z^{D-2}}\sum_{k_{D-1}=1}^{\infty}\left(
\sum_{j=1}^{D-1}\left(\frac{2\pi k_j}{L_j}\right)^2\right)^{-\alpha
s}\\+&\sum_{(k_1,\ldots, k_{D-2})\in
\Z^{D-2}\setminus\{\mathbf{0}\}}\left(
\sum_{j=1}^{D-2}\left(\frac{2\pi k_j}{L_j}\right)^2\right)^{-\alpha
s}=Y_1(s)+Y_2(s).\nonumber
\end{align*}  In the
limit $L_i\rightarrow \infty$ for all $1\leq i\leq D-1$,
\begin{align*}
Y_1(s)=&2\sum_{(k_1,\ldots, k_{D-2})\in
\Z^{D-2}}\sum_{k_{D-1}=1}^{\infty}\left(
\sum_{j=1}^{D-1}\left(\frac{2\pi k_j}{L_j}\right)^2\right)^{-\alpha
s}\\\nonumber=&\frac{2L_1\ldots
L_{D-2}}{(2\pi)^{D-2}}\sum_{k_{D-1}=1}^{\infty}\int\limits_{\R^{D-2}}d^{D-2}\mathbf{k}\left(|\mathbf{k}|^2+
\left(\frac{2\pi k_{D-1}}{L_{D-1}}\right)^2\right)^{-\alpha s}\\
\nonumber =&\frac{4\pi^{(D-2)/2}L_1\ldots
L_{D-2}}{(2\pi)^{D-2}\Gamma\left(\frac{D-2}{2}\right)}\sum_{k_{D-1}=1}^{\infty}\int_0^{\infty}
k^{D-3}\left(k^2+ \left(\frac{2\pi
k_{D-1}}{L_{D-1}}\right)^2\right)^{-\alpha s}dk\\
\nonumber =&\frac{2\pi^{(D-2)/2}L_1\ldots
L_{D-2}}{(2\pi)^{D-2}}\frac{\Gamma\left(\alpha
s-\frac{D-2}{2}\right)}{\Gamma(\alpha s)}\sum_{k_{D-1}=1}^{\infty}
\left(\frac{2\pi
k_{D-1}}{L_{D-1}}\right)^{D-2-2\alpha s}\\
\nonumber=&\frac{2\pi^{(D-2)/2}L_1\ldots
L_{D-2}}{(2\pi)^{D-2}}\left(\frac{2\pi
}{L_{D-1}}\right)^{D-2-2\alpha s}\frac{\Gamma\left(\alpha
s-\frac{D-2}{2}\right)}{\Gamma(\alpha s)}\zeta_R(2\alpha s-(D-2))\\
\nonumber=&\frac{\pi^{2\alpha s -\frac{3(D-2)+1}{2}}L_1\ldots
L_{D-2}}{2^{D-3}\Gamma(\alpha s)}\left(\frac{2\pi
}{L_{D-1}}\right)^{D-2-2\alpha s}\Gamma\left(\frac{D-1}{2}-\alpha
s\right)\zeta_R(D-1-2\alpha s).
\end{align*}Therefore,
\begin{align*}
Y_1'(0)=\frac{\pi^{ -\frac{3(D-2)+1}{2}}L_1\ldots
L_{D-2}}{2^{D-3}\Gamma(\alpha s)}\left(\frac{2\pi
}{L_{D-1}}\right)^{D-2}\Gamma\left(\frac{D-1}{2}\right)\zeta_R(D-1)
\end{align*}and the limit $\lim_{L_{i}\rightarrow
\infty}\bigl(Y_1'(0)/(L_1\ldots L_{D-1})\bigr)$ vanishes. Similarly,
the limit $\lim_{L_{i}\rightarrow \infty}\bigl(Y_2'(0)/(L_1\ldots
L_{D-1})\bigr)=0$. Consequently, the contribution to the free energy
density only comes from $\zeta_{\alpha; \chi,\chi}^1\!'(0)$ and we
find that the free energy density and the pressure in this case are
given respectively by \eqref{eq207} and \eqref{eq208} by putting
$\eta=0$. In particular, when $D=3$,
\begin{align}\label{eq106}
f_{\alpha;\chi,\chi}=&-\frac{\alpha}{d^3}\left(
\frac{\pi^6\xi^4}{45}+\frac{\pi^2}{720
}+4\pi^2\xi^4\sum_{m=1}^{\infty}\sum_{n=1}^{\infty}
\frac{1}{(m^2+(2\pi\xi n)^2)^{2}}
\right),\end{align}\begin{align*}
P_{\alpha;\chi,\chi}=&\frac{\alpha}{d^4}\left(
\frac{\pi^6\xi^4}{45}-\frac{\pi^2}{240
}+4\pi^2\xi^4\sum_{m=1}^{\infty}\sum_{n=1}^{\infty} \frac{m^2-3(2\pi
\xi n)^2}{(m^2+(2\pi\xi n)^2)^{3}} \right).
\end{align*}

 \vspace{0.3cm}
\subsection{The case $\chi=\mu= 1$ [Neumann Boundary Condition]}~

 \vspace{0.2cm}\noindent
In this case, the corresponding zeta function \eqref{eq205} is given
by
\begin{align*} \zeta_{\alpha;
1,1}=\sum_{n=0}^{\infty}\sum_{m\in\Z}\sum_{\mathbf{k}\in
\Z^{D-1}}\!'\left( \sum_{j=1}^{D-1}\left(\frac{2\pi
k_j}{L_j}\right)^2+\left(\frac{\pi n}{d}\right)^2+\left(\frac{2\pi
m}{\beta}\right)^2\right)^{-\alpha s}.
\end{align*}It is easy to see that the sum of $\zeta_{\alpha; 1,1}$ with  $\zeta_{\alpha; 0, 0}$ \eqref{eq7}
gives $\zeta_{\alpha;\chi,\chi}, \chi\neq 0, 1$ \eqref{eq298}.
Therefore \begin{align*} \zeta_{\alpha;
1,1}'(0)=\zeta_{\alpha;\chi,\chi}'(0)-\zeta_{\alpha; 0,0}'(0).
\end{align*} We obtain from \eqref{eq14_1} in Section \ref{sec2} and \eqref{eq207}
in Section \ref{sec1} (with $\eta=0$) that in this case, the free
energy density is given by\begin{align*} f_{\alpha;
1,1}=&-\frac{\alpha\Gamma\left(\frac{D+1}{2}\right)\pi^{\frac{D+1}{2}}}{d^D}\Biggl(
\xi^{D+1}\zeta_R(D+1)+\frac{\zeta_R(D+1)}{(2
\pi)^{D+1}}+\frac{\Gamma\left(\frac{D}{2}\right)}
{2\sqrt{\pi}\Gamma\left(\frac{D+1}{2}\right)}\xi^D\zeta_R(D)\\
&\hspace{3cm}+2\xi^{D+1}\sum_{m=1}^{\infty}\sum_{n=1}^{\infty}
\frac{1}{(m^2+(2\pi \xi n)^2)^{(D+1)/2}}\nonumber\Biggr),
\end{align*}and the pressure $P_{\alpha; 1,1}=P_{\alpha;0,0}$.
In particular, when $D=3$,
\begin{align*}
f_{\alpha; 1, 1}=-\frac{\alpha}{2d^3} \left(
\frac{\pi^6\xi^4}{45}+\frac{\pi^2}{720
}+4\pi^2\xi^4\sum_{m=1}^{\infty}\sum_{n=1}^{\infty}
\frac{1}{(m^2+(2\pi\xi n)^2)^{2}}
+\frac{\pi^2\xi^3}{2}\zeta_R(3)\right).
\end{align*}

 \vspace{0.3cm}\subsection{The case $(\chi,\mu)=(1,0)$ or $(0,1)$ [Boyer Boundary Condition]
}~

 \vspace{0.2cm}\noindent
In this case, the corresponding zeta function \eqref{eq205} becomes
\begin{align*}
\zeta_{\alpha;
0,1}(s)=\sum_{n=1}^{\infty}\sum_{m\in\Z}\sum_{\mathbf{k}\in
\Z^{D-1}}\left( \sum_{j=1}^{D-1}\left(\frac{2\pi
k_j}{L_j}\right)^2+\left(\frac{\pi
}{d}\left(n-\frac{1}{2}\right)\right)^2+\left(\frac{2\pi
m}{\beta}\right)^2\right)^{-\alpha s}.
\end{align*}Observe that
\begin{align*}
\zeta_{\alpha; 0,
1}(s)=\frac{1}{2}\sum_{n\in\Z}\sum_{m\in\Z}\sum_{\mathbf{k}\in
\Z^{D-1}}\left( \sum_{j=1}^{D-1}\left(\frac{2\pi
k_j}{L_j}\right)^2+\left(\frac{\pi
}{d}\left(n-\frac{1}{2}\right)\right)^2+\left(\frac{2\pi
m}{\beta}\right)^2\right)^{-\alpha s}.
\end{align*}Therefore, we can obtain the free energy density for this case by multiplying
\eqref{eq207} in Section \ref{sec1} by $1/2$ and setting $\eta=1$.
This gives us
\begin{align*}
f_{\alpha;
0,1}=&-\frac{\alpha\Gamma\left(\frac{D+1}{2}\right)\pi^{\frac{D+1}{2}}}{d^D}\Biggl(
\xi^{D+1}\zeta_R(D+1)-\frac{1-2^{-D}}{(2 \pi)^{D+1}}\zeta_R(D+1)\\
&\hspace{3cm}+2\xi^{D+1}\sum_{m=1}^{\infty}\sum_{n=1}^{\infty}
\frac{(-1)^n}{(m^2+(2\pi \xi n)^2)^{(D+1)/2}}\nonumber\Biggr),
\end{align*}and
\begin{align*}
P_{\alpha;0,1}=&\frac{\alpha\Gamma\left(\frac{D+1}{2}\right)\pi^{\frac{D+1}{2}}}{d^{D+1}}\Biggl(
\xi^{D+1}\zeta_R(D+1)-\frac{D(1-2^{-D})}{(2 \pi)^{D+1}}\zeta_R(D+1)\\
&\hspace{3cm}+2\xi^{D+1}\sum_{m=1}^{\infty}\sum_{n=1}^{\infty}
\frac{(-1)^n(m^2-D(2\pi \xi n)^2)}{(m^2+(2\pi \xi
n)^2)^{(D+3)/2}}\nonumber\Biggr).
\end{align*}When $D=3$,
\begin{align}\label{eq40}
f_{\alpha;0,1}=-\frac{\alpha}{2d^3}  \left(
\frac{\pi^6\xi^4}{45}-\frac{7}{8}\frac{\pi^2}{720
}+4\pi^2\xi^4\sum_{m=1}^{\infty}\sum_{n=1}^{\infty}
\frac{(-1)^n}{(m^2+(2\pi\xi n)^2)^{2}} \right),
\end{align}\begin{align*}
P_{\alpha; 0,1}=\frac{\alpha}{2d^4}  \left(
\frac{\pi^6\xi^4}{45}+\frac{7}{8}\frac{\pi^2}{240
d^4}+4\pi^2\xi^4\sum_{m=1}^{\infty}\sum_{n=1}^{\infty}
\frac{(-1)^n(m^2-3(2\pi\xi n)^2)}{(m^2+(2\pi\xi n)^2)^{3}} \right).
\end{align*}

\vspace{1cm} The results obtained for various boundary conditions
can now be summarized in the following compact form:
\begin{align}\label{eq220} f_{\alpha;
\chi,\mu}=&-\frac{\sigma_{\chi,\mu}\alpha\Gamma\left(\frac{D+1}{2}\right)\pi^{\frac{D+1}{2}}}{d^D}\Biggl(
\xi^{D+1}\zeta_R(D+1)+\frac{1}{(2 \pi)^{D+1}}\sum_{n=1}^{\infty}
\frac{\cos(\pi n \eta)}{n^{D+1}}\\
&+2\xi^{D+1}\sum_{m=1}^{\infty}\sum_{n=1}^{\infty} \frac{\cos(\pi n
\eta)}{(m^2+(2\pi \xi
n)^2)^{(D+1)/2}}+\omega_{\chi,\mu}\frac{\Gamma\left(\frac{D}{2}\right)}
{2\sqrt{\pi}\Gamma\left(\frac{D+1}{2}\right)}\xi^D\zeta_R(D)\nonumber\Biggr),
\end{align}
\begin{align*}
P_{\alpha;\chi,\mu}=&\frac{\sigma_{\chi,\mu}\alpha\Gamma\left(\frac{D+1}{2}\right)\pi^{\frac{D+1}{2}}}{d^{D+1}}\Biggl(
\xi^{D+1}\zeta_R(D+1)-\frac{D}{(2 \pi)^{D+1}}\sum_{n=1}^{\infty}
\frac{\cos(\pi n \eta)}{n^{D+1}}\\
&\hspace{3cm}+2\xi^{D+1}\sum_{m=1}^{\infty}\sum_{n=1}^{\infty}
\frac{\cos(\pi n \eta)(m^2-D(2\pi \xi n)^2)}{(m^2+(2\pi \xi
n)^2)^{(D+3)/2}}\nonumber\Biggr).
\end{align*}where
\begin{align*}
\sigma_{\chi,\mu}=&\begin{cases} 1,
\hspace{1cm}\;&\text{if}\;\;(\chi,\mu)=(0,0), (0,1), (1,0),
(1,1);\\
2,&\text{else}.\end{cases}\\\omega_{\chi,\mu}=&\begin{cases} 1,
\hspace{1cm}\;&\text{if}\;\;(\chi,\mu)=
(1,1);\\
-1, &\text{if}\;\;(\chi,\mu)=(0,0);\\ 0,&\text{else}.\end{cases}
\end{align*}

Tracking back the derivation of formula \eqref{eq220}, we can also
write the free energy density as (see \eqref{eq15})
\begin{align}\label{eq310}
f_{\alpha; \chi,\mu}= -\frac{\sigma_{\chi,\mu}\alpha
d\Gamma\left(\frac{D+1}{2}\right)}{2^{D+2}\pi^{\frac{3(D+1)}{2}}}Z_E\left(
\frac{D+1}{2};\frac{1}{a_1},\frac{1}{a_2}; 0 ,
-\mathbf{g}\right)-\sigma_{\chi,\mu}\omega_{\chi,\mu}\frac{\alpha
\pi^{\frac{D}{2}}\Gamma\left(\frac{D}{2}\right)}{2d^D}
\xi^D\zeta_R(D),
\end{align}with $\mathbf{g}=(-\eta/2, 0)$.

 \vspace{0.3cm}
\subsection{Casimir Energy of Electromagnetic field confined between
parallel walls}\label{section4}~

As is well known (see e.g. \cite{AW}), the Casimir energy of
electromagnetic field in four dimensional space--time confined
between two infinite parallel plates can be computed using almost
the same setup as the massless scalar field with $D=3$ and
$\alpha=1$. More specifically, since there are two transverse
polarization for electromagnetic fields, its free energy will be
twice that of the massless scalar field. In the case when the two
parallel plates are both perfectly conduction, except for the factor
of two, it is almost equivalent to the Dirichlet boundary condition.
However, as pointed out in \cite{B, AW, F}, an additional $1/2$ of
the $n=0$ modes must be added. This amounts to the omission of the
second term in \eqref{eq20}. Therefore,  the Casimir energy density
for the fractional electromagnetic field is given by
\begin{align}\label{eq210}
f_{Cas}=-\frac{\alpha}{d^3} \left(
\frac{\pi^6\xi^4}{45}+\frac{\pi^2}{720
}+4\pi^2\xi^4\sum_{m=1}^{\infty}\sum_{n=1}^{\infty}
\frac{1}{(m^2+(2\pi\xi n)^2)^{2}} \right),
\end{align}in perfect agreement with the result obtained in \cite{BM} when $\alpha=1$.
In Boyer's setup, where one plate is perfectly conduction and the
other is infinitely permeable, the result for electromagnetic field
should be twice the result for massless scalar field under Boyer's
boundary condition. In fact, when $D=3$ and $\alpha=1$, twice of the
formula \eqref{eq40} agree with the result obtained in \cite{n33}.

By these comparisons with electromagnetic field, one can provide a
heuristic interpretation regarding the fractional Neumann boundary
conditions \eqref{eq1} imposed on the parallel plates as their
deviation from the perfect conductivity and infinite permeability.

\section{\textbf{Low and High Temperature Expansion and Limit of the Free Energy Density}}In this section,
we consider the low and high temperature limits of the free energy
density. For this purpose, a generalization of the Chowla--Selberg
formula for Epstein zeta function (see e.g. \cite{E3, E4}) is
particularly useful. We have
\begin{align*}
&Z_E(s; c_1, c_2; 0;\mathbf{h}) -2\sum_{n=1}^{\infty}
\frac{\cos(2\pi n h_1)}{(c_1
n^2)^s}\\=&\frac{2}{\Gamma(s)}\sum_{m=1}^{\infty}\cos(2\pi m
h_2)\int_0^{\infty} t^{s-1}\sum_{n=-\infty}^{\infty}
e^{-t(c_1n^2+c_2m^2)+2\pi i n h_1}dt\\
=&\frac{2\sqrt{\pi}}{\sqrt{c_1}\;\Gamma(s)}\sum_{m=1}^{\infty}\cos(2\pi
m h_2)\int_0^{\infty}
t^{s-(3/2)}\sum_{n=-\infty}^{\infty}e^{-t(c_2m^2)-\frac{\pi^2}{tc_1}(n-h_1)^2}dt\nonumber.
\end{align*}Here we have used the Poisson summation formula. If $h_1=0$, then we have to separate the $n=0$ term and
obtain
\begin{align}\label{eq320}
&Z_E(s; c_1, c_2;
0;\mathbf{h})=2c_1^{-s}\zeta_R(2s)+\frac{2\sqrt{\pi}\;\Gamma\left(s-\frac{1}{2}\right)}{c_2^{s-(1/2)}\sqrt{c_1}\;\Gamma(s)}\sum_{m=1}^{\infty}
\frac{\cos(2\pi m h_2)}{m^{2s-1}}\\
&+\frac{8\sqrt{\pi}}{\sqrt{c_1}\;\Gamma(s)}\sum_{n=1}^{\infty}\sum_{m=1}^{\infty}\cos(2\pi
m h_2)\left(\frac{\pi
n}{\sqrt{c_1c_2}m}\right)^{s-(1/2)}K_{s-(1/2)}\left(2\pi\sqrt{\frac{c_2}{c_1}}mn\right).\nonumber
\end{align}If $0<h_1<1$, then
\begin{align}\label{eq321}
&Z_E(s; c_1, c_2; 0;\mathbf{h})=2c_1^{-s}\sum_{n=1}^{\infty}
\frac{\cos(2\pi n
h_1)}{n^{2s}}\\&+\frac{4\sqrt{\pi}}{\sqrt{c_1}\;\Gamma(s)}\sum_{n=-\infty}^{\infty}\sum_{m=1}^{\infty}\cos(2\pi
m h_2)\left(\frac{\pi
|n-h_1|}{\sqrt{c_1c_2}m}\right)^{s-(1/2)}K_{s-(1/2)}\left(2\pi\sqrt{\frac{c_2}{c_1}}m|n-h_1|\right).\nonumber
\end{align}

 \vspace{0.2cm}
\subsection{Low Temperature Expansion}~
 \vspace{0.1cm}

\noindent By taking $c_1=1/a_1, c_2=1/a_2,\mathbf{h}=(\eta/2, 0)$ in
\eqref{eq320} and \eqref{eq321} , we have the low temperature
 ($T\ll 1$ or $\xi\ll 1$) expansion of the free energy density \eqref{eq310},
 i.e. when $\eta=0$,
\begin{align*}
f_{\alpha; \chi,\mu}=& -\frac{\sigma_{\chi,\mu}\alpha
d\Gamma\left(\frac{D+1}{2}\right)}{2^{D+2}\pi^{\frac{3(D+1)}{2}}}\Biggl(\frac{2\pi^{D+1}}{d^{D+1}}
\zeta_R(D+1)+(1+\omega_{\chi,\mu})\frac{2^{D+1}\pi^{2D+(3/2)}\Gamma\left(\frac{D}{2}\right)}{d^{D+1}\Gamma\left(\frac{D+1}
{2}\right)}\xi^{D}\zeta_R(D)\\
&+\frac{2^{(D/2)+3}\pi^{2D+(3/2)}\xi^{D/2}}{d^{D+1}\Gamma\left(\frac{D+1}{2}\right)}
\sum_{m=1}^{\infty}\sum_{n=1}^{\infty}\left(\frac{n}{m}\right)^{D/2}K_{D/2}
\left(\frac{nm}{\xi}\right)\Biggr),
\end{align*}and when $\eta\neq 0$,
\begin{align*}
f_{\alpha; \chi,\mu}=& -\frac{\sigma_{\chi,\mu}\alpha
d\Gamma\left(\frac{D+1}{2}\right)}{2^{D+2}\pi^{\frac{3(D+1)}{2}}}\Biggl(2\left(\frac{\pi}{d}\right)^{D+1}
\sum_{m=1}^{\infty}\frac{\cos(\pi
n\eta)}{n^{D+1}}\\
&+\frac{2^{(D/2)+2}\pi^{2D+(3/2)}\xi^{D/2}}{d^{D+1}\Gamma\left(\frac{D+1}{2}\right)}
\sum_{m=1}^{\infty}\sum_{n=-\infty}^{\infty}\left(\frac{1}{m}\left|n-\frac{\eta}{2}\right|\right)^{D/2}K_{D/2}
\left(\left|n-\frac{\eta}{2}\right|\frac{m}{\xi}\right)\Biggr).
\end{align*}From \cite{AAR}, pg 223 , we have the following asymptotic
expansion for $K_{D/2}(z)$:
\begin{align}\label{eq322}
K_{\nu}(z)\sim
\sqrt{\frac{\pi}{2z}}e^{-z}\left(1+\sum_{k=1}^{\infty}\frac{(\nu,k)}{(2z)^k}\right),
\end{align}where
\begin{align*}
(\nu,
k)=\frac{1}{2^{2k}k!}\prod_{i=1}^{k}\bigl(4\nu^2-(2i-1)^2\bigr).
\end{align*}When $\nu$ is equal to half of an odd integer, the sum in \eqref{eq322} is finite and the right hand side
of \eqref{eq322} is the exact formula for $K_{\nu}(z)$. From the
asymptotic expansion \eqref{eq322}, we see that when $\xi\rightarrow
0$,
\begin{align*}
\sum_{m=1}^{\infty}\sum_{n=1}^{\infty}\left(\frac{n}{m}\right)^{D/2}K_{D/2}
\left(\frac{nm}{\xi}\right)
\end{align*}is exponentially decay and the leading term is obtained
by setting $m=n=1$, which results in
\begin{align*}
\sum_{m=1}^{\infty}\sum_{n=1}^{\infty}\left(\frac{n}{m}\right)^{D/2}K_{D/2}
\left(\frac{nm}{\xi}\right)\sim \sqrt{\frac{\pi
\xi}{2}}\left(1+\sum_{k=1}^{\infty}\frac{(D/2,
k)}{2^k}\xi^{k}\right)e^{-\frac{1}{\xi}}+O\left(e^{-\frac{2}{\xi}}\right).
\end{align*}Similarly, when $0<\eta<1$, the $m=1,n=0$ term gives
\begin{align*}
&\sum_{m=1}^{\infty}\sum_{n=-\infty}^{\infty}\left(\frac{1}{m}\left|n-\frac{\eta}{2}\right|\right)^{D/2}K_{D/2}
\left(\left|n-\frac{\eta}{2}\right|\frac{m}{\xi}\right)\Biggr)\\&\sim\left(\frac{\eta}{2}\right)^{(D-1)/2}
\sqrt{\frac{\pi \xi}{2}}\left(1+\sum_{k=1}^{\infty}\frac{(D/2,
k)}{\eta^k}\xi^{k}\right)
e^{-\frac{\eta}{2\xi}}+O\left(e^{-\frac{\min\{\eta,1-(\eta/2)\}}{\xi}}\right).
\end{align*}When $\eta=1$, the $m=1,n=\pm 1$ terms give
\begin{align*}
&\sum_{m=1}^{\infty}\sum_{n=-\infty}^{\infty}\left(\frac{1}{m}\left|n-\frac{\eta}{2}\right|\right)^{D/2}K_{D/2}
\left(\left|n-\frac{\eta}{2}\right|\frac{m}{\xi}\right)\Biggr)\\&\sim
\frac{1}{2^{(D-2)/2}} \sqrt{\pi \xi}\left(1+\sum_{k=1}^{\infty}(D/2,
k)\xi^{k}\right)
e^{-\frac{1}{2\xi}}+O\left(e^{-\frac{1}{\xi}}\right).
\end{align*}
These imply that for low temperature $T\ll 1$, when $\eta=0$,
\begin{align}\label{eq401} f_{\alpha; \chi,\mu}=&
-\frac{\sigma_{\chi,\mu}\alpha }{d^D}\Biggl(\frac{
\Gamma\left(\frac{D+1}{2}\right)}{2^{D+1}\pi^{(D+1)/2}}
\zeta_R(D+1)+(1+\omega_{\chi,\mu})\frac{\pi^{D/2}\Gamma\left(\frac{D}{2}\right)}{2}\xi^D\zeta_R(D)\\
&+\frac{(\pi\xi)^{(D+1)/2}}{2^{(D-1)/2}}\left(1+\sum_{k=1}^{\infty}\frac{(D/2,
k)}{2^k}\xi^{k}\right)e^{-\frac{1}{\xi}}\Biggr)+O\left(e^{-\frac{2}{\xi}}\right),\nonumber
\end{align}when $0<\eta< 1$,
\begin{align}\label{eq15_1}
f_{\alpha; \chi,\mu}=&-\frac{\sigma_{\chi,\mu}\alpha
}{d^D}\Biggl(\frac{
\Gamma\left(\frac{D+1}{2}\right)}{2^{D+1}\pi^{(D+1)/2}}
\sum_{m=1}^{\infty}\frac{\cos(\pi
n\eta)}{n^{D+1}}\\&+\frac{(\pi\xi)^{(D+1)/2}\eta^{(D-1)/2}}{2^D}\left(1+\sum_{k=1}^{\infty}\frac{(D/2,
k)}{\eta^k}\xi^{k}\right)
e^{-\frac{\eta}{2\xi}}\Biggr)+O\left(e^{-\frac{\min\{\eta,1-(\eta/2)\}}{\xi}}\right),\nonumber
\end{align} and finally when $\eta=1$,\begin{align}\label{eq403}
f_{\alpha; \chi,\mu}=&-\frac{\sigma_{\chi,\mu}\alpha
}{d^D}\Biggl(-\frac{
\Gamma\left(\frac{D+1}{2}\right)}{2^{D+1}\pi^{(D+1)/2}}(1-2^{-D})
\zeta_R(D+1)\\&+\frac{(\pi\xi)^{(D+1)/2}}{2^{D-1}}\left(1+\sum_{k=1}^{\infty}(D/2,
k)\xi^{k}\right)
e^{-\frac{1}{2\xi}}\Biggr)+O\left(e^{-\frac{1}{\xi}}\right)\nonumber.
\end{align} From these, we also find that the zero
temperature energy density is
\begin{align*}
f_{\alpha; \chi,\mu}^0=&-\frac{\sigma_{\chi,\mu}\alpha }{d^D}\frac{
\Gamma\left(\frac{D+1}{2}\right)}{2^{D+1}\pi^{(D+1)/2}}
\sum_{m=1}^{\infty}\frac{\cos(\pi n\eta)}{n^{D+1}}.
\end{align*}This term depends on $\eta$. For $D=2, 3, 4, 5$, the relation between the
normalized zero temperature energy density $d^Df_{\alpha; 0,\eta}^0$
and $\eta$ is shown in Figure 1.
\begin{figure}\centering \epsfxsize=.75\linewidth
\epsffile{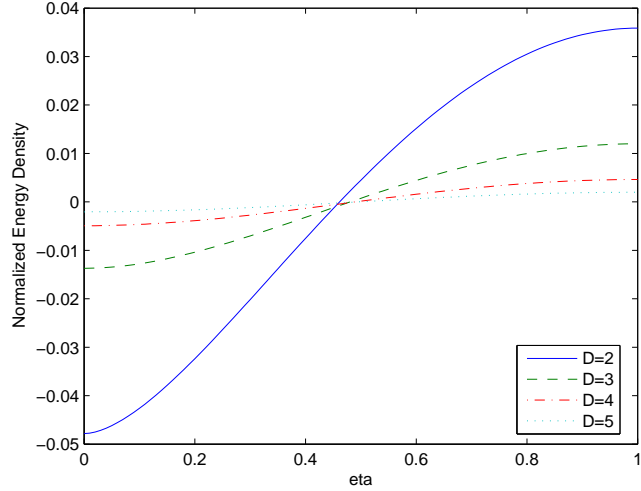} \label{Figure1}\caption{The normalized
zero temperature free energy density $d^Df_{\alpha; 0,\eta}^0$ for
$D=2,3,4,5$ and $\alpha=1$. The horizontal axis is the $\eta$ axis.}
\end{figure}
 When $\eta=0$, its value
\begin{align*}
f_{\alpha; \chi,\mu}^0=&-\frac{\sigma_{\chi,\mu}\alpha }{d^D}\frac{
\Gamma\left(\frac{D+1}{2}\right)}{2^{D+1}\pi^{(D+1)/2}}\zeta_R(D+1)
\end{align*}is negative, and when $\eta=1$, its value
\begin{align*}
f_{\alpha; \chi,\mu}^0=&-\frac{\sigma_{\chi,\mu}\alpha }{d^D}\frac{
\Gamma\left(\frac{D+1}{2}\right)}{2^{D+1}\pi^{(D+1)/2}}\sum_{n=1}^{\infty}\frac{(-1)^n}{n^{D+1}}\\=&
(1-2^{-D})\frac{\sigma_{\chi,\mu}\alpha }{d^D}\frac{
\Gamma\left(\frac{D+1}{2}\right)}{2^{D+1}\pi^{(D+1)/2}} \zeta_R(D+1)
\end{align*}is positive. We are going to show in the Appendix that
the function
\begin{align}\label{eq404}
\mathfrak{B}_n(x)=\sum_{k=1}^{\infty}\frac{\cos(2\pi k x)}{k^n},
\hspace{1cm}0\leq x\leq 1, \;\;n\geq 2
\end{align}is increasing in the interval $[0, 1/2]$. Consequently,
when $\eta$ changes from $0$ to $1$, the zero temperature energy
density increases, and it changes from negative to positive, so that
the nature of the force in the system changes accordingly from
attractive to repulsive. For a specific $D$, there is a transition
value $\eta_D$ so that the force is attractive when $\eta\in[0,
\eta_D)$ and the force is repulsive when $\eta\in(\eta_D, 1]$. We
tabulate some values of
$\eta_D$ in Table 6.1.\\

 Table 6.1\\
\begin{tabular}{c|c|c|c}
\hline\hspace{0.3cm} $D$\hspace{0.3cm} &\hspace{1cm} $\eta_D$ \hspace{1cm} & \hspace{0.3cm} $D$ \hspace{0.3cm} &
\hspace{0.6cm} $\eta_D$\hspace{1.2cm}\\
 \hline
2 & 0.4226  & 3 & \hspace{1cm}0.4617\hspace{1cm} \\
4 & 0.4807 & 5 & \hspace{1cm}0.4902\hspace{1cm} \\
6 & 0.4951 & 7 & \hspace{1cm}0.4975\hspace{1cm} \\
\hline
\end{tabular}\\
\\

\vspace{0.2cm}

From this table, we see that $\eta_D$ is an increasing function of
$D$. We have verified numerically that this is true for all $D\leq
45$. On the other hand, we can in fact show mathematically that
$\eta_D<0.5$ for all $D$ (see Appendix). In Figure 2, we show the
graph of $\eta_D$ as a function of $D$.

\begin{figure}\centering \epsfxsize=.6\linewidth
\epsffile{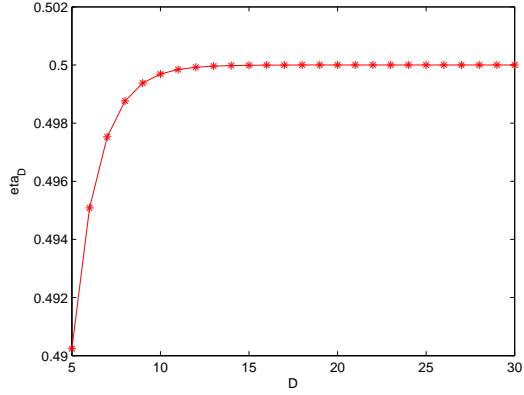} \label{Figure10}\caption{$\eta_D$ (eta$_D$) as a
function of $D$.}
\end{figure}

From \eqref{eq401}, \eqref{eq15_1} and \eqref{eq403}, we also find
that when $\chi\neq \mu$ or $(\chi,\mu)=(0,0)$, the thermal
correction to the zero temperature energy decays exponentially,
whereas if $\chi=\mu\neq 0$, there is a term proportional to $T^D$.

When $\xi\ll 1$, the dependence of the normalized free energy
density on $\xi$ and $\eta=\mu-\chi$ for $D=3$ is shown in Figure 3
and Figure 4.

\begin{figure}\centering \epsfxsize=.6\linewidth
\epsffile{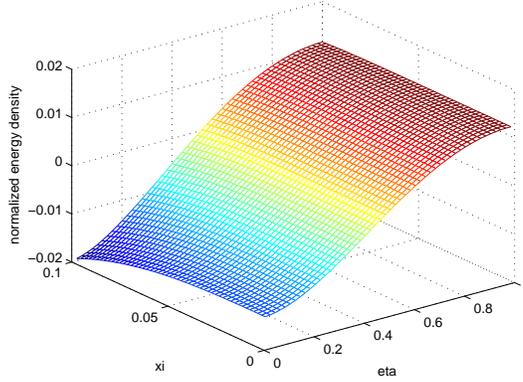} \label{Figure2}\caption{The normalized free
energy density $d^Df_{\alpha; 0,\eta}$ for $D=3$ and $\alpha=1$. The
$x$ and $y$ axes are the $\xi$ (xi) and  $\eta$ (eta) axes.}
\end{figure}

\begin{figure}\centering \epsfxsize=.6\linewidth
\epsffile{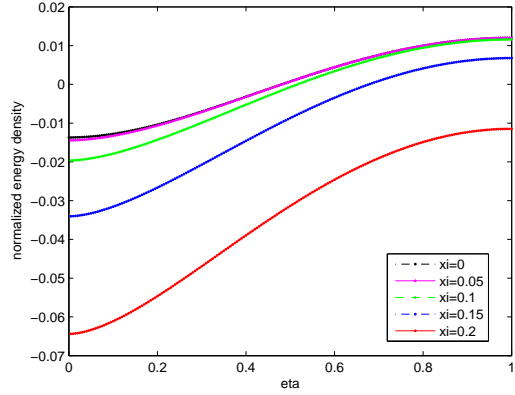} \label{Figure2}\caption{The normalized free
energy density $d^Df_{\alpha; 0,\eta}$ for $D=3$ and $\alpha=1$ when
$\xi=0, 0.05, 0.1, 0.15, 0.2$ respectively.}
\end{figure}

 \vspace{0.3cm}
\subsection{High Temperature Expansion}~

 \vspace{0.1cm}Take $c_1=1/a_2, c_2=1/a_1,\mathbf{h}=(0, \eta/2)$ in \eqref{eq320},
we have the high temperature ($T\gg 1$ or $\xi\gg 1$)  expansion of
the free energy density \eqref{eq310}, .i.e. \begin{align*}
f_{\alpha; \chi,\mu}=& -\frac{\sigma_{\chi,\mu}\alpha
d\Gamma\left(\frac{D+1}{2}\right)}{2^{D+2}\pi^{\frac{3(D+1)}{2}}}\Biggl(2\left(\frac{2\pi^2\xi}{d}\right)^{D+1}
\zeta_R(D+1)+\frac{4\pi^{D+(5/2)}\Gamma\left(\frac{D}{2}\right)}{d^{D+1}\Gamma\left(\frac{D+1}{2}\right)}\xi\sum_{n=1}^{\infty}
\frac{\cos(\pi n \eta)}{n^{D}}\\
&+\frac{2^{(D/2)+4}\pi^{2D+(5/2)}}{d^{D+1}\Gamma\left(\frac{D+1}{2}\right)}\xi^{(D+2)/2}\sum_{n=1}^{\infty}\sum_{m=1}^{\infty}\cos(\pi
n\eta)\left(\frac{
m}{n}\right)^{D/2}K_{D/2}\left(4\pi^2mn\xi\right)\Biggr).
\end{align*}Using the asymptotic expansion of the modified Bessel function
\eqref{eq322}, we find that if $T\gg 1$ (or equivalently $\xi\gg
1$),
\begin{align}\label{eq402}
 f_{\alpha;\chi,\mu}\sim &-\frac{\sigma_{\chi,\mu}\alpha
}{d^D}\Biggl(\pi^{(D+1)/2}\Gamma\left(\frac{D+1}{2}\right)\zeta_R(D+1)\xi^{D+1}+\frac{\Gamma\left(\frac{D}{2}\right)}
{2^D\pi^{(D/2)-1}}\xi\sum_{n=1}^{\infty} \frac{\cos(\pi n
\eta)}{n^{D}}\\&+\frac{\pi^{(D+1)/2}}{2^{(D-1)/2}}\cos(\pi\eta)\xi^{(D+1)/2}\left(1+\sum_{k=1}^{\infty}
\frac{(D/2,k)}{(8\pi^2\xi)^k}\right)e^{-4\pi^2\xi}\Biggr)+O\left(e^{-8\pi^2\xi}\right).\nonumber
\end{align}The leading term
\begin{align*}
f_{\alpha;\chi,\mu}^{\infty, 1}=-\frac{\sigma_{\chi,\mu}\alpha
}{d^D}\pi^{(D+1)/2}\Gamma\left(\frac{D+1}{2}\right)\zeta_R(D+1)\xi^{D+1}
\end{align*}is proportional to $T^{D+1}$, and is independent of
$\eta$. When $D=3$, it gives
\begin{align*}
f_{\alpha;\chi,\mu}^{\infty,
1}=-\sigma_{\chi,\mu}\alpha\frac{\pi^2d}{90}T^4,
\end{align*}which is called the Stefan-Boltzmann term. The next
leading term of the energy density at high temperature is
proportional to $T$, with proportionality constant depends on
$\eta$. The rest of the terms decay exponentially. From this, we can
conclude that when the temperature is large enough, the effect of
different boundary conditions is not significant and the system
exhibits a universal behavior regardless of the boundary conditions.
When $\xi\gg 1$, the dependence of the normalized free energy
density on $\xi$ and $\eta=\mu-\chi$ for $D=3$ is shown in Figure 5.

\begin{figure}\centering \epsfxsize=.65\linewidth
\epsffile{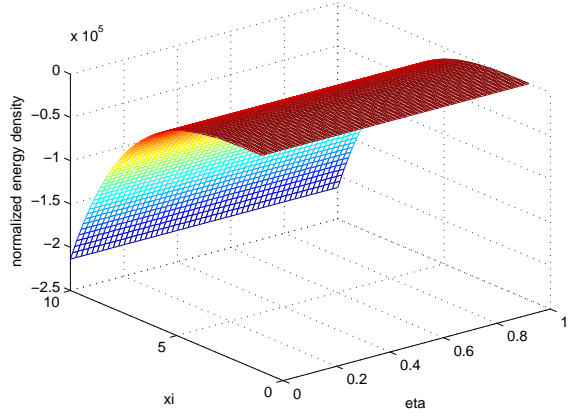} \label{Figure3}\caption{The normalized free
energy density $d^Df_{\alpha; 0,\eta}$ for $D=3$ and $\alpha=1$. The
$x$ and $y$ axes are the   $\eta$ (eta) and $\xi$ (xi) axes.}
\end{figure}

As we explain in Section \ref{section4}, if we take
$\sigma_{\chi,\mu}=2$, $\omega_{\chi,\mu}=0$, $\alpha=1$, $\eta=0$
and $D=3$ in the energy density $f_{\alpha; \chi,\mu}$
\eqref{eq310}, we obtain the Casimir energy for electromagnetic
fields confined between perfectly conducting parallel  infinite
plates \eqref{eq210}. Therefore, we obtain from \eqref{eq401} and
\eqref{eq402} the low and high temperature limit of the Casimir
energy \eqref{eq210}:
\begin{align*}
f_{Cas}\sim& -\frac{1 }{d^3}\Biggl(\frac{\pi^2}{720}+
\frac{\pi^{2}}{2}\zeta_R(3)\xi^3+\pi^2\left(\xi^2+\xi^3\right)e^{-\frac{1}{\xi}}\Biggr)+O\left(e^{-\frac{2}{\xi}}\right),
\hspace{0.3cm}\xi\ll 1,\\
f_{Cas}\sim & -\frac{1 }{d^3}\Biggl(\frac{\pi^6}{45}\xi^4+\frac{\xi}
{8}\zeta_R(3)+\left(\pi^2\xi^2+\frac{\xi}{4}\right)e^{-4\pi^2\xi}\Biggr)+O\left(e^{-8\pi^2\xi}\right),\hspace{0.2cm}\xi\gg
1,
\end{align*}agree with the result of \cite{BM}. On the other hand,
if we take $\sigma_{\chi,\mu}=2$, $\omega_{\chi,\mu}=0$, $\alpha=1$,
$\eta=1$ and $D=3$, we obtain the Casimir energy for electromagnetic
fields confined between one perfectly conducting and one infinitely
permeable parallel  infinite plates. Therefore, from \eqref{eq403}
and \eqref{eq402}, we find that the low and high temperature limits
of the Casimir energy density of this system are
\begin{align*}
-\frac{1
}{d^3}\Biggl(-\frac{7\pi^2}{5760}+\pi^2\left(\frac{\xi^2}{2}+\xi^3\right)e^{-\frac{1}{2\xi}}\Biggr)+O\left(e^{-\frac{1}{\xi}}\right),
\hspace{1cm}\xi\ll 1,\\
-\frac{1}{d^3}\Biggl(\frac{\pi^6}{45}\xi^4-\frac{3}{32}\zeta_R(3)\xi-\left(\pi^2\xi^2+\frac{\xi}{4}\right)e^{-4\pi^2\xi}\Biggr)
+O\left(e^{-8\pi^2\xi}\right), \hspace{0.3cm}\xi\gg 1.
\end{align*}These agree with the results in \cite{n33}.

\section{\textbf{Temperature Inversion Symmetry}}Since the observation of the
symmetry between low and high temperature exhibited by the Casimir
energy between perfectly conduction parallel plates \eqref{eq210}
pointed out by Brown and Maclay in \cite{BM}, there have been a
number of papers devoted to the discussion of the temperature
inversion symmetry of different systems \cite{n33, n34, n35, n36,
n37, LR, PBPR}. Here we want to point out the mathematical origin of
this symmetry, and show that in some particular cases, the free
energy density \eqref{eq220} exhibits temperature inversion
symmetry.

Observe that when $\mathbf{h}=0=\mathbf{g}$, the Epstein zeta
function \eqref{eq11} is completely symmetric with respect to $a_1$
and $a_2$. In particular, if we define
\begin{align*}
H_s(w)=\sum_{(n_1,n_2)\in\Z^2}\!'
\frac{1}{\left(n_1^2+(n_2w)^2\right)^s}=Z_E(s; 1, w^2;\mathbf{0},
\mathbf{0} ),
\end{align*}then
\begin{align*}
Z_E(s; a_1, a_2; \mathbf{0}, \mathbf{0})=
a_1^{-s}H_s\left(\sqrt{\frac{a_2}{a_1}}\right)=a_2^{-s}H_s\left(\sqrt{\frac{a_1}{a_2}}\right).
\end{align*}The symmetry of the Epstein zeta
function expressed using $H_s$ is the relation
\begin{align}\label{eq221}
H_s(w) = w^{-2s}H_s\left(\frac{1}{w}\right).
\end{align}In our case, $\sqrt{a_2/a_1}=2\pi \xi=2dT$ and formulas
in the form \eqref{eq221} precisely gives a relation between low and
high temperature.

 Using the formula \eqref{eq310} for free energy density, when $\chi=\mu$,
we have $\mathbf{g}=\mathbf{0}$ and therefore
\begin{align*}
d^Df_{\alpha; \chi,\chi}=&-\frac{\sigma_{\chi,\mu}\alpha
\Gamma\left(\frac{D+1}{2}\right)}{2^{D+2}\pi^{\frac{3(D+1)}{2}}}d^{D+1}a_2^{\frac{D+1}{2}}
H_{\frac{D+1}{2}}\left(\sqrt{\frac{a_2}{a_1}}\right)-
\sigma_{\chi,\mu}\omega_{\chi,\mu}\frac{\alpha
\pi^{\frac{D}{2}}\Gamma\left(\frac{D}{2}\right)}{2}
\xi^D\zeta_R(D).\end{align*}The second term is zero except when
$\chi=\mu=0$ or 1. The first term, denoted by $\mathcal{F}_0$, is a
function of $\xi$ and is equal to
\begin{align*}\mathcal{F}_0(\xi)=&- \frac{\sigma_{\chi,\mu}\alpha
\Gamma\left(\frac{D+1}{2}\right)\pi^{\frac{D+1}{2}}}{2}\xi^{D+1}
H_{\frac{D+1}{2}}\left(2\pi\xi\right).
\end{align*}From \eqref{eq221}, it satisfies the inversion symmetry
\begin{align*}
\mathcal{F}_0(\xi)=(2\pi\xi)^{D+1}\mathcal{F}_0\left(\frac{1}{4\pi^2\xi}\right).
\end{align*}For $D=3$, $\alpha=1$, this is precisely the symmetry observed in
\cite{BM, n36} for electromagnetic field confined between parallel
perfectly conducting plates. Therefore, when $\chi=\mu\neq 0,1$, the
normalized free energy density of a massless fractional Klein-Gordon
field confined between two parallel hyperplanes
$d^Df_{\alpha;\chi,\mu}$ has a complete temperature inversion
symmetry. When $\chi=\mu=0,1$, the symmetry is broken by a term
proportional to $\xi^D\zeta_R(D)$.

When $\eta=\pm 1$ or equivalently, $(\chi,\mu)=(0,1)$ or $(1,0)$,
from \eqref{eq310} we have\begin{align*} d^Df_{\alpha;
0,1}=&-\frac{\alpha
\Gamma\left(\frac{D+1}{2}\right)}{2^{D+2}\pi^{\frac{3(D+1)}{2}}}(\sqrt{a_1a_2}d)^{D+1}\sum_{m\in\Z}
\sum_{n\in\Z}\!'\frac{(-1)^n}{(a_1 m^2+a_2n^2)^{(D+1)/2}}.
\end{align*}
We can rewrite the double sum as \begin{align*}& \sum_{m\in\Z}
\sum_{n\in\Z}\!'\frac{(-1)^n}{(a_1
m^2+a_2n^2)^{(D+1)/2}}\\=&2\sum_{m\in\Z}
\sum_{n\in\Z}\!'\frac{1}{(a_1
m^2+a_2(2n)^2)^{(D+1)/2}}-\sum_{m\in\Z}
\sum_{n\in\Z}\!'\frac{1}{(a_1 m^2+a_2n^2)^{(D+1)/2}}.
\end{align*}Therefore, the normalized free energy density $ d^Df_{\alpha; 0,1}$ can be written as a sum of two functions in $\xi$,
$_1\!\mathcal{F}_1(\xi)$ and $_2\!\mathcal{F}_1(\xi)$ where
\begin{align*}
_1\!\mathcal{F}_1(\xi)=&-\alpha
\Gamma\left(\frac{D+1}{2}\right)\pi^{\frac{D+1}{2}}\xi^{D+1}
H_{\frac{D+1}{2}}\left(4\pi\xi\right),\\
_2\!\mathcal{F}_1(\xi)=&\frac{\alpha}{2}
\Gamma\left(\frac{D+1}{2}\right)\pi^{\frac{D+1}{2}}\xi^{D+1}
H_{\frac{D+1}{2}}\left(2\pi\xi\right).
\end{align*}Using \eqref{eq221}, we find that each of these
functions
satisfies an inversion symmetry
\begin{align*}
_1\!\mathcal{F}_1(\xi)=(4\pi\xi)^{D+1}\;_1\!\mathcal{F}_1\left(\frac{1}{16\pi^2\xi}\right),\hspace{1cm}
_2\!\mathcal{F}_1(\xi)=(2\pi\xi)^{D+1}\;_2\!\mathcal{F}_1\left(\frac{1}{4\pi^2\xi}\right).
\end{align*}When $D=3$, $\alpha=1$,
this is what observed in \cite{n33} for electromagnetic field confined between parallel plates under
Boyer's setup.

For generic $\eta$, there was no temperature inversion symmetry
since the components in $\mathbf{g}$ are not symmetric. However, for
some particular rational values of $\eta$, we can use the same trick
as in the case $\eta=1$ and write the normalized energy density as a
sum of a few functions such that each of them has temperature
inversion symmetry. For example, when $\eta=2/3$, using the fact
that when $f(x)$ is an even function,
\begin{align*}
\sum_{n\in\Z}e^{\pi i n\eta} f(n)=&f(0)+2\Biggl(\sum_{n\geq 1,
n\equiv 0 \;\text{mod}\; 3}f(n)+\sum_{n\geq 1, n\equiv
1\;\text{mod}\;
3}\cos\left(\frac{2\pi}{3}\right)f(n)\\&\hspace{3cm}+\sum_{n\geq 1,
n\equiv 2 \;\text{mod}\;
3}\cos\left(\frac{4\pi}{3}\right)f(n)\Biggr)\\
=&f(0)+3\sum_{n\geq 1, n\equiv 0 \;\text{mod}\;
3}f(n)-\sum_{n=1}^{\infty}f(n)\\
=&\frac{3}{2}\sum_{n\in\Z}f(3n)-\frac{1}{2}\sum_{n\in\Z}f(n),
\end{align*}we can write $d^Df_{\alpha; \chi, \chi\pm (2/3)}$ as a
sum of two functions $_1\!\mathcal{F}_2(\xi)$,
$_2\!\mathcal{F}_2(\xi)$ given by
\begin{align*}
_1\!\mathcal{F}_2(\xi)=&-\frac{3}{2}\alpha
\Gamma\left(\frac{D+1}{2}\right)\pi^{\frac{D+1}{2}}\xi^{D+1}
H_{\frac{D+1}{2}}\left(6\pi\xi\right),\\
_2\!\mathcal{F}_2(\xi)=&\frac{\alpha}{2}
\Gamma\left(\frac{D+1}{2}\right)\pi^{\frac{D+1}{2}}\xi^{D+1}
H_{\frac{D+1}{2}}\left(2\pi\xi\right).
\end{align*}Each of these functions
satisfies temperature inversion symmetry
\begin{align*}
_1\!\mathcal{F}_2(\xi)=(6\pi\xi)^{D+1}\;_1\!\mathcal{F}_2\left(\frac{1}{36\pi^2\xi}\right),\hspace{1cm}
_2\!\mathcal{F}_2(\xi)=(2\pi\xi)^{D+1}\;_2\!\mathcal{F}_2\left(\frac{1}{4\pi^2\xi}\right).
\end{align*}

\section{\textbf{Conclusion}}
We have introduced a new type of boundary condition called
fractional Neumann condition which involves vanishing fractional
derivative of the field in the study of Casimir effect of fractional
massless Klein-Gordon field confined between a pair of parallel
plates. By imposing  this fractional Neumann conditions on the
plates allows the interpolation between the usual Dirichlet and
Neumann conditions. Our results indicate that there exists a
transition value for the difference between the orders of the
fractional Neumann conditions in the two plates for which the
Casimir force changes from attractive to repulsive (or vice versa).
It is interesting to note that for sufficiently high temperature,
the Hemholtz free energy density is dominated by a term independent
of boundary conditions. Conditions for temperature inversion
symmetry to hold are also discussed.

 We would also
like to point out that despite a few decades of work on temperature
dependence of Casimir effect, there still exist debates on this
topic. The main issue of the recent controversy lies in the
thermodynamic consistency of the computed Casimir force between real
metals and the Drude dispersion relation (see references
\cite{r1,r2,r3,r4,r5} for both sides of the controversy). It has to
do with the controversy of inclusion/exclusion of the TE (transverse
electric) zero mode. Some authors \cite{r6,r7,r8} claimed that the
Drude relation does not provide a consistent explanation of recent
experimental results, in particular it is in conflict with the
Nernst theorem. They proposed to replace the Drude relation by the
plasma relation. On the other hand, Hoye, Brevik, Aarseth, Ellingsen
and Milton \cite{r2, r4, r9, r10, r11, r12, r13, r14, r18} have
argued in favor of the exclusion of the TE zero mode. They have
derived analytical results using Euler-Maclaurin formula, which in
the limit $T\rightarrow 0$ are consistent with the Nernst theorem
\cite{r17}. They have also carried out numerical calculation of the
free energy and obtained results which agree with analytic results
to a high degree of accuracy. These authors also proposed an
experimental setup to test such results \cite{r18}. We plan to
discuss in detail the Casimir energy of fractional electromagnetic
field and the issue of inclusion/exclusion of the TE zero mode in a
future work.

    Finally, we would like to suggest some other possible directions for further work.
The extension of our discussion to a $p$--dimensional cavity
embedded in $D$--dimensional space, with $p\leq D$ is currently
under consideration. However for the generalization of the above
results to non-flat space is likely to encounter highly non-trivial
mathematical problems since one needs to deal with fractional
operators in curved space. Another interesting generalization
involves fractional Klein-Gordon field with fractional Neumann
boundary conditions of variable fractional order, which allows
variable Casimir energy or force at different point in space. Such a
problem again requires results from derivatives and integrations of
fractional variable order, a subject which is still at its infancy.

\vspace{1cm} \noindent \textbf{Acknowledgement}\; S.C. Lim and L.P.
Teo would like to thank Malaysian Academy of Sciences, Ministry of
Science, Technology and Innovation for funding this project under
the Scientific Advancement Fund Allocation (SAGA) Ref. No P96c.

\appendix \section{\textbf{The Function $\mathfrak{B}_n(x)$}} In this appendix, we are going to show that the function
\eqref{eq404}
\begin{align*}
\mathfrak{B}_n(x)=\sum_{k=1}^{\infty}\frac{\cos(2\pi k x)}{k^n}
\end{align*}is increasing and has exactly one zero in the interval $[0,
1/2]$. We are also going to show that this unique zero is less than
$1/4$.

First, we show that $\mathfrak{B}_n(x)$ is increasing and has
exactly one zero in the interval $[0, 1/2]$. As a matter of fact,
for $n$ even, the function $\mathfrak{B}_n(x)$ is well known. From
\textbf{9.622} of \cite{GR}, we have
\begin{align}\label{eq411}
\mathfrak{B}_{2n}(x)=(-1)^{n-1}\frac{(2\pi)^{2n}}{2(2n)!}B_{2n}(x),
\hspace{1cm}0\leq x\leq 1,
\end{align}where $B_{k}(x)$ is the $k$-th Bernoulli polynomial
defined by
\begin{align*}
\frac{te^{tx}}{e^t-1}=\sum_{k=0}^{\infty}\frac{B_k(x)}{k!}t^k.
\end{align*}The explicit formula for $B_k(x)$ for $1\leq k\leq 5$ is given in Table A.1.\\

\hspace{2cm} Table A.1\\
\begin{tabular}[A.1]
{c|c}
\hline \hspace{0.3cm} $k$ \hspace{0.3cm} &  \hspace{0.3cm}$B_k(x)$\\
\hline 1 &  \hspace{0.3cm}$x-\frac{1}{2}$\\
2 &  \hspace{0.3cm}$x^2-x+\frac{1}{6}$\\
3 & \hspace{0.3cm} $x^3-\frac{3}{2}x^2+\frac{1}{2}x$\\
4 & $x^4-2x^3+x^2-\frac{1}{30}$\\
5 & $ x^5-\frac{5}{2}x^4+\frac{5}{3}x^3-\frac{1}{6}x$\\
\hline
\end{tabular}\\

 \vspace{0.2cm}
It is well known that for all $k\geq 1$, $B_{k+1}'(x)= (k+1)B_k(x)$.
From \textbf{9.622} of \cite{GR} again, we have
\begin{align}\label{eq412} B_{2n-1}(x)=\frac{(-1)^n 2
(2n-1)!}{(2\pi)^{2n-1}}\sum_{k=1}^{\infty}\frac{\sin (2k\pi
x)}{k^{2n-1}}, \hspace{1cm} \begin{cases}k=1,  & 0<x<1 \\k\geq 2, &
0\leq x\leq 1 \end{cases}.
\end{align}From this it is easy to verify that $B_{2n}(x)$ is
increasing and has exactly one zero in the interval $[0, 1/2]$ (see
e.g. \cite{E}). For convenience, we repeat the argument here. As is
easily verify from \eqref{eq411},
\begin{align*}
B_{2n}(0)=&2(-1)^{n-1}\frac{(2n)!}{(2\pi)^{2n}}\zeta_R(2n),\\
B_{2n}(1/2)=&2(1-2^{1-2n})(-1)^{n}\frac{(2n)!}{(2\pi)^{2n}}\zeta_R(2n),
\end{align*}which shows that $B_{2n}(0)$ and $B_{2n}(1/2)$ has
opposite sign and is nonzero. It also implies that $B_{2n}(x)$ must
has at least one zero in $[0,1/2]$. On the other hand, we find from
\eqref{eq412} that $B_{2k+1}(0)=B_{2k+1}(1/2)=0$ for all $k\geq 1$.
Now if for some $j\geq 1$, $B_{2j}(x)$ has two zeros in $[0, 1/2]$,
then its derivative $2j B_{2j-1}(x)$ has a zero in $(0, 1/2)$. Since
$B_{2j-1}(0)=B_{2j-1}(1/2)=0$, this in turn implies that its
derivative $(2j-1)B_{2j-2}(x)$ has two zeros in $[0, 1/2]$.
Continuing this argument, we find that $B_1(x)$ must have a zero in
$(0, 1/2)$. This gives a contradiction since $B_1(x)=x-(1/2)$ does
not have any zero in $(0,1/2)$. This shows that $B_{2n}(x)$ has
exactly one zero in the interval $[0, 1/2]$ and $B_{2n-1}(x)$ does
not have any zero in the open interval $(0, 1/2)$. The latter
implies that $B_{2n-1}(x)$ must be either always nonnegative or
always nonpositive in the interval $(0, 1/2)$. Therefore,
$B_{2n}(x)$ is monotone in $[0, 1/2]$. This completes our argument
for $\mathfrak{B}_{2n}(x)$.

To verify the statement for $\mathfrak{B}_{2n-1}(x)$, $n\geq 1$, we
define the functions
\begin{align*}
\mathfrak{D}_{n}(x)=\sum_{k=1}^{\infty}\frac{\sin(2k\pi x)}{k^n},
\hspace{1cm} \begin{cases}n=1,  & 0<x<1 \\n\geq 2, & 0\leq x\leq 1
\end{cases}
\end{align*}and let
\begin{align*}
C_{2n-1}(x) =\frac{
(-1)^{n-1}}{(2\pi)^{2n-1}}\mathfrak{B}_{2n-1}(x), \hspace{1cm}
C_{2n}(x)=\frac{(-1)^{n-1}}{(2\pi)^{2n}}\mathfrak{D}_{2n}(x)
\end{align*}for all $n\geq 1$. Then it is easy to verify that
$C_{n+1}'(x)=C_{n}(x)$. Moreover, for all $n\geq 1$,
\begin{align*}
C_{2n+1}(0)=&\frac{(-1)^n}{(2\pi)^{2n+1}}\zeta_R(2n+1),\\
C_{2n+1}(1/2)=&-(1-2^{-2n})\frac{(-1)^{n}}{(2\pi)^{2n+1}}\zeta_R(2n+1),
\end{align*}$C_{2n}(0)=C_{2n}(1/2)=0$. On the other hand, for
$0<x<1$,
\begin{align*}
\mathfrak{B}_1(x)+i\mathfrak{D}_1(x)=\sum_{k=1}^{\infty}
\frac{e^{2\pi i k x}}{k}=-\log\left(1-e^{2\pi i x}\right)=-\log
(2\sin(\pi x))-i\pi\left(x-\frac{1}{2}\right).
\end{align*}Therefore, $$C_1(x)=\frac{1}{2\pi}\mathfrak{B}_1(x)=-\frac{1}{2\pi}\log(2\sin(\pi x))$$and
it is easy to verify that $C_1(x)$ is decreasing on $(0, 1/2)$,
positive on $(0, 1/6)$, negative on $(1/6, 1/2]$ and zero at
$x=1/6$. Since $C_2'(x)=C_1(x)$, we find that $C_2$ is strictly
increasing on $[0, 1/6]$ and strictly decreasing on $[1/6, 1/2]$.
Since $C_2(0)=C_2(1/2)=0$, $C_2(x)>0$ for all $x\in (0, 1/2)$. The
same argument used for $B_{2n}$ then shows that $C_{2n+1}(x)$ is
monotone and has exactly one zero in the interval $[0,1/2]$, thus
verifying the statement for $\mathfrak{B}_{2n-1}(x)$.

Now since $\mathfrak{B}_{n}(0)=\zeta_R(n)>0$,
$\mathfrak{B}_{n}(1/2)=-(1-2^{1-n})\zeta_R(n)<0$, to show that the
unique zero of $\mathfrak{B}_{n}(x)$ is less than $1/4$, it is
enough to show that $\mathfrak{B}_n(1/4)<0$. A straightforward
computation gives
\begin{align*}
\mathfrak{B}_n\left(\frac{1}{4}\right)=
-\frac{1}{2^n}+\frac{1}{4^n}-\frac{1}{6^n}+\frac{1}{8^n}+\ldots=\frac{1}{2^n}\mathfrak{B}_n\left(\frac{1}{2}\right)<0,
\end{align*}verifying our claim.


\begin{thebibliography}{10}
\bibitem{n1}
R. Hilfer (Ed.), \emph{Applications of Fractional Calculus in
Physics}, World Scientific,
     Singapore, (2000).


\bibitem{n2} G.~M. Zaslavsky, \emph{Chaos, fractional kinetics, and anomalous transport},
Phys. Rep. \textbf{371}  (2002), 461--580.

\bibitem{n3}
 B.~J. West, M. Bologna and P. Grigolini, \emph{Physics of Fractal
Operators}, Springer--Verlag, New York, (2003).


\bibitem{n4}
R. Metzler and J. Klafter, \emph{The restaurant at the end of the
random walk: recent developments in the description of anomalous
transport by fractional dynamics}, J. Phys. A \textbf{37} (2004),
R161--R208.

\bibitem{n5}  L.~M. Zelenyi and A.~V. Milovanov, \emph{Fractal topology and strange kinetics: from percolation theory to problems
in cosmic electrodynamics}, Phys. Uspekhi \textbf{47} (2004), no.~8, 749--788.


\bibitem{n6} G.~M. Zaslavsky, \emph{Hamiltonian Chaos and Fractional Dynamics}, Oxford University,
     Oxford, (2005).


\bibitem{n7} N. Laskin, \emph{Fractional quantum mechanics}, Phys. Rev. E \textbf{62} (2000), 3135--3145.

\bibitem{n8} N. Laskin, \emph{Fractals and quantum mechanics}, Chaos \textbf{10} (2000), 780--790.

\bibitem{n9} N. Laskin, \emph{Fractional {S}chrodinger equation}, Phys. Rev. E \textbf{66} (2002), Art. No. 056108.

\bibitem{n10} M. Naber, \emph{Time fractional {S}chrodinger equation}, J. Math. Phys. \textbf{45} (2004), 3339--3352.

\bibitem{n11} X. Guo and M. XU, \emph{Some physical applications of fractional {S}chrodinger
equation}, J. Math. Phys. \textbf{47} (2006), Art. No. 082104.

\bibitem{n12}
C.~G. Bollini and J.~J. Giambiagi, \emph{Arbitrary powers of
d'{A}lembertians
  and the {H}uygens principle}, J. Math. Phys. \textbf{34} (1993), no.~2,
  610--621.


\bibitem{n13}
Claus L{\"a}mmerzahl, \emph{The pseudodifferential operator square
root of the
  {K}lein-{G}ordon equation}, J. Math. Phys. \textbf{34} (1993), no.~9,
  3918--3932.

\bibitem{n14} S. Albeverio H. Gottschalk and J.-L Wu,
\emph{Convoluted generalized white noise, Schwinger functions and
their analytic continuation to Wightman functions}, Rev. Math. Phys.
\textbf{8} (1996), 763--817.

\bibitem{n15} M. Grothaus and L. Streit, \emph{Construction of relativistic quantum fields in the framework of white noise analysis},
J. Math. Phys. \textbf{40} (1999), 5387--5405.

\bibitem{n16}
M.~S. Plyushchay and Michel Rausch~de Traubenberg, \emph{Cubic root
of
  {K}lein-{G}ordon equation}, Phys. Lett. B \textbf{477} (2000), no.~1-3,
  276--284.


\bibitem{n17} A. Raspini, \emph{Simple solutions of the fractional Dirac equation of order 2/3},
Physica Scripta \textbf{64} (2001), 20--22.

\bibitem{n18}
Petr Z{\'a}vada, \emph{Relativistic wave equations with fractional
derivatives
  and pseudodifferential operators}, J. Appl. Math. \textbf{2} (2002), no.~4,
  163--197.

\bibitem{n19}
R.~L. P.~G. do~Amaral and E.~C. Marino, \emph{Canonical quantization
of
  theories containing fractional powers of the d'{A}lembertian operator}, J.
  Phys. A \textbf{25} (1992), no.~19, 5183--5200.

\bibitem{n20}
D.~G. Barci, L.~E. Oxman, and M.~Rocca, \emph{Canonical quantization
of
  nonlocal field equations}, Internat. J. Modern Phys. A \textbf{11} (1996),
  no.~12, 2111--2126.

\bibitem{n21}
S.~C. Lim and S.~V. Muniandy, \emph{Stochastic quantization of
nonlocal
  fields}, Phys. Lett. A \textbf{324} (2004), no.~5-6, 396--405.

\bibitem{n22}
S.~C. Lim, \emph{Fractional derivative quantum fields at positive
temperature},
  Phys. A \textbf{363} (2006), no.~2, 269--281.



\bibitem{n23} A.G. Riess et al.
[Supernova Search Team Collaboration], \emph{Observational evidence
from supernovae for an accelerating universe and a cosmological
constant}, Astron. J. \textbf{116} (1998), 1009--1038.

\bibitem{n24}  S. Perlmutter et al. [Suernova Cosmology Project
Collaboration], \emph{Measurements of {O}mega and {L}ambda from 42
high-redshift supernovae}, Astrophys. J. \textbf{517} (1999),
565--586.

\bibitem{n25}  J.L. Tonry et al. [Supernova Search Team
Collaboration], \emph{Cosmological results from high-z supernovae},
Astrophys. J. \textbf{594} (2003), 1--24.

\bibitem{n26}  A.G. Riess et al. [Supernova Search Team
Collaboration], \emph{Type {I}a supernova discoveries at $z > 1$
from the {H}ubble {S}pace {T}elescope: Evidence for past
deceleration and constraints on dark energy evolution}, Astrophys.
J. \textbf{607} (2004), 665--687.

\bibitem{n27}  P.W. Milonni, \emph{The Quantum Vacuum}, Academic Press, New York,
(1994).

\bibitem{n28}  K.A. Milton, \emph{The Casimir Effect}, World-Scientific, Singapore,
(2001)

\bibitem{n29}  K.A. Milton, \emph{Dark Energy as Evidence for Extra Dimensions}, Grav. Cosmol. \textbf{9} (2003),
66--70.

\bibitem{n30} P. Brax, J. Martin,
J-P Uzan, Eds. \emph{On the Nature of Dark Energy}, Proceedings of
18th IAP Astrophysics Colloquium, Frontier Group, (2002).

\bibitem{n31} T.H. Boyer, \emph{Van der {W}aals forces and zero-point energy for dielectric and permeable materials,}
 Phys. Rev. A \textbf{9} (1974), 2078--2084.

\bibitem{n32} M.V. Cougo-Pinto, C. Farina, J.F.M. Mendes and A.C. Tort,
\emph{zeta-function method for repulsive Casimir forces},  Braz. J.
Phys. \textbf{29} (1999), 371--374.

\bibitem{n33}
F.~C. Santos, A. Tenorio and  A.~C. Tort,  \emph{Zeta function
method and repulsive
  {C}asimir forces for an unusual pair of plates at finite temperature},
  Physical Review D \textbf{60} (1999), no.~10, Art. No. 105022.

\bibitem{n34} S. Tadaki and S. Takagi, \emph{Casimir Effect at Finite Temperature},
 Prog. Theor. Phys. \textbf{75} (1982), 262--271.

\bibitem{n35} S.A. Gundersen and F. Ravndal, \emph{The {F}ermionic {C}asimir effect at finite temperature},
Ann. Phys. (N.Y.) \textbf{182}  (1988), 90--111.

\bibitem{n36}
F.~Ravndal and D.~Tollefsen, \emph{Temperature inversion symmetry in
the
  {C}asimir effect}, Physical Review D \textbf{40} (1989), no.~12, 4191--4192.


\bibitem{n37}
\bysame, \emph{On the {C}asimir effect and the temperature inversion
symmetry},
  J. Phys. A \textbf{23} (1990), no.~9, 1627--1632.


\bibitem{LR}
C.~Wotzasek, \emph{A symmetry in the finite-temperature {C}asimir
effect}, J.
  Phys. A \textbf{21} (1988), L793--L796.






\bibitem{PBPR}
A.~C.~Aguiar Pinto, T.~M. Britto, F.~Pascoal, and F.~S.~S. da~Rosa,
  \emph{Temperature inversion symmetry in the {C}asimir effect with an
  antiperiodic boundary condition}, Phys. Rev. D \textbf{67} (2003), Art. No.
  107701.

\bibitem{n38}
Hitoshi Kumano-go, \emph{Pseudodifferential operators}, MIT Press,
Cambridge,
  Mass., (1981).


\bibitem{n39} S. Samko, A.A. Kilbas and D.I. Maritchev, \emph{Integrals and
Derivatives of the Fractional Order and Some of Their Applications},
Gordon and Breach, Armsterdam, (1993).


\bibitem{n40}
C.~G. Bollini and J.J. Giambiagi, \emph{Lagrangian procedures for
higher order
  field equations}, Revista Brasileira de F$\acute{\text{i}}$sica \textbf{17}
  (1987), no.~1, 14--30.


\bibitem{n41}
A.~Pais and G.~E. Uhlenbeck, \emph{On field theories with
non-localized
  action}, Physical Rev. (2) \textbf{79} (1950), 145--165.

\bibitem{n42}
E.~C. Marino, \emph{Complete bosonization of the {D}irac fermion
field in
 {$2+1$} dimensions}, Phys. Lett. B \textbf{263} (1991), no.~1, 63--68.

\bibitem{n43}
D.G. Barci, C{\'e}sar~D. Fosco, and L.E. Oxman, \emph{On
bosonization in $3$ dimensions},
 Phys. Lett.
\textbf{B375}
  (1996), no.~1, 267--272.

\bibitem{n44}
A.~O. Barvinsky and G.~A. Vilkovisky, \emph{Beyond the
{S}chwinger-{D}e{W}itt technique: Converting loops into trees and
in-in currents}, Nucl. Phys.  \textbf{B282} (1987), 163--188;
\emph{Covariant perturbation theory.
  {II}.\ {S}econd order in the curvature. {G}eneral algorithms}, Nucl. Phys.
   \textbf{B333} (1990), no.~2, 471--511.

\bibitem{n45}
Diego A.~R. Dalvit and Francisco~D. Mazzitelli, \emph{Running
coupling
  constants, {N}ewtonian potential, and nonlocalities in the effective action},
  Phys. Rev. D \textbf{50} (1994), no.~2, 1001--1009.

\bibitem{n46}  L. Nottale, \emph{Fractal Space-Time and Microphysics}, World
Scientific, Singapore, (1993).

\bibitem{n47} H. Kroger, \emph{Fractal geometry in quantum mechanics, field theory and spin systems},
Phys. Rep. \textbf{323} (2000), 82--181.

\bibitem{n48} J.S. Dowker and R. Critchley, \emph{Effective
{L}agrangian and energy-momentum tensor in de{S}itter space}, Phys.
Rev. \textbf{D13} (1976), 3224--3232.

\bibitem{n49}
S.W. Hawking, \emph{Zeta function regularization of path integrals
in curved spacetime}, Commun. Math. Phys. \textbf{55} (1977),
133--148.

\bibitem{n50} G.W. Gibbons, \emph{Thermal zeta functions}, Phys. Lett. \textbf{A60} (1977), 385--386.

\bibitem{AW}
Jan Ambj{\o}rn and S.~Wolfram, \emph{Properties of the vacuum. {I}.
  {M}echanical and thermodynamic}, Ann. Physics \textbf{147} (1983), 1--32.

\bibitem{E1}
E.~Elizalde, S.~D. Odintsov, A.~Romeo, A.~A. Bytsenko, and
S.~Zerbini,
  \emph{Zeta regularization techniques with applications}, World Scientific
  Publishing Co. Inc., River Edge, NJ, 1994.

\bibitem{E2}
Emilio Elizalde, \emph{Ten physical applications of spectral zeta
functions},
  Lecture Notes in Physics. New Series m: Monographs, vol.~35, Springer-Verlag,
  Berlin, 1995.



\bibitem{K}
K.~Kirsten, \emph{Spectral functions in mathematics and physics},
Chapman \&
  Hall/ CRC, Boca Raton, FL, 2002.

 \bibitem{GR}
I.~S. Gradshteyn and I.~M. Ryzhik, \emph{Table of integrals, series,
and
  products}, sixth ed., Academic Press Inc., San Diego, CA, 2000, Translated
  from the Russian.


\bibitem{B}
G.~Barton, \emph{Quantum electrodynamics of spinless particles
between
  conducting plates}, Proc. R. Soc. Lond. \textbf{320} (1970), no.~1541,
  251--275.

\bibitem{F}
C.~Farina, \emph{Casimir effect: some aspects}, Braz. J. Phys.
\textbf{36}
  (2006), 1137--1149.

\bibitem{BM}
Lowell~S. Brown and G.~Jordan Maclay, \emph{Vacuum stress between
conduction
  plates: an image solution}, Physical Review \textbf{184} (1969), no.~5,
  1272--1279.

\bibitem{E3}
E.~Elizalde, \emph{An extension of the {C}howla-{S}elberg formula
useful in
  quantizing with the {W}heeler-{D}e{W}itt equation}, J. Phys. A \textbf{27}
  (1994), no.~11, 3775--3785.

\bibitem{E4}
\bysame, \emph{Zeta functions: formulas and applications}, J.
Comput. Appl.
  Math. \textbf{118} (2000), no.~1-2, 125--142, Higher transcendental functions
  and their applications.


 \bibitem{AAR}
George~E. Andrews, Richard Askey, and Ranjan Roy, \emph{Special
functions},
  Encyclopedia of Mathematics and its Applications, vol.~71, Cambridge
  University Press, Cambridge, 1999.




















\bibitem{r1}
R.~S. Decca, D. Lopez, E. Fischbach et al,  \emph{Precise comparison
of theory and new experiment for the {C}asimir force leads to
stronger constraints on thermal quantum effects and long-range
interactions}, Ann. Phys. \textbf{318} (2005), no.~1, 37--80.



\bibitem{r2}
I. Brevik, J.~B. Aarseth, J.~S. H{\o}ye et al,  \emph{Temperature
dependence of the {C}asimir effect}, Phys. Rev. E \textbf{71}
(2005), no.~5, Art. No. 056101.

\bibitem{r3}
V.~B. Bezerra, R.~S. Decca, E. Fischbach et al, \emph{Comment on
"{T}emperature dependence of the {C}asimir effect"}, Phys. Rev. E
\textbf{73} (2006), no.~2, Art. No. 028101.

\bibitem{r4}
J.~S. H{\o}ye, I. Brevik, J.~B. Aarseth et al, \emph{What is the
temperature dependence of the {C}asimir effect?} J.Phys. A
\textbf{39} (2006), no.~20, 6031--6038.


\bibitem{r5}
V.~M. Mostepanenko, V.~B. Bezerra, R. S. Decca et al, \emph{Present
status of controversies regarding the thermal {C}asimir force}, J.
Phys. A \textbf{39} (2006), no.~21,  6589--6600.

\bibitem{r6}
G.~L. Klimchitskaya, V.~M. Mostepanenko, \emph{Investigation of the
temperature dependence of the {C}asimir force between real metals},
Phys. Rev. A \textbf{63} (2001), no.~6, Art. No. 062108.

\bibitem{r7}
M. Bordag, B. Geyer B, G.~L. Klimchitskaya et al, \emph{Casimir
force at both nonzero temperature and finite conductivity}, Phys.
Rev. Lett. \textbf{85}, (2000), no.~3, 503--506.


\bibitem{r8}
E. Fischbach, D.~E. Krause, V.~M. Mostepanenko et al, \emph{New
constraints on ultrashort-ranged {Y}ukawa interactions from atomic
force microscopy}, Phys. Rev. D \textbf{64} (2001), no.~7, Art. No.
075010.




\bibitem{r9}
J.~S. H{\o}ye, I. Brevik, J.~B. Aarseth et al, \emph{Does the
transverse electric zero mode contribute to the Casimir effect for a
metal?} Phys. Rev E \textbf{67} (2003), no. 5, Art. No. 056116.

\bibitem{r10}
M. Bostrom, B.~E. Sernelius, \emph{Thermal effects on the {C}asimir
force in the 0.1--5 $\mu$m range}, Phys. Rev. Lett. \textbf{84}
(2000), no.~20, 4757--4760.

\bibitem{r11}
I. Brevik, J.~B. Aarseth, \emph{Temperature dependence of the
Casimir effect}, J. Phys. A \textbf{39} (2006), no.~21, 6187--6193.


\bibitem{r12}
I. Brevik, J. B. Aarseth, J. S. Hoye, and K. A. Milton,
\emph{Temperature dependence of the {C}asimir force for metals}, in
\emph{Qunatum field theory under the influence of external
conditions}, edited by K. A. Milton, Rinton Press, Princeton, NJ,
2004, 54--65.

\bibitem{r13}
I. Brevik, S.~A. Ellingsen and K.~A. Milton, \emph{Thermal
corrections to the Casimir effect},  New J.  Phys. \textbf{8}
(2006),  Art. No. 236.


\bibitem{r14}
S.~A. Ellingsen, \emph{Casimir attraction in multilayered plane
parallel magnetodielectric systems}, J. Phys. A \textbf{40}, no.~9,
1951--1961.

\bibitem{r18}
S.~A. Ellingsen and I. Brevik, \emph{Casimir force on real materials
- the slab and cavity geometry}, J. Phys. A \textbf{40} (2007),
no.~13, 3643--3664.



\bibitem{r17}
J.~S. H{\o}ye, I. Brevik, S.~A. Ellingsen and J.~B. Aarseth et al,
\emph{Analytical and numerical verification of the {N}ernst theorem
for metals}, preprint arXiv:quant-ph/0703174 (2007).



\bibitem{E}
H.~M. Edwards, \emph{Riemann's zeta function}, Dover Publications
Inc.,
  Mineola, NY, (2001).





\end{thebibliography}
\end{document}